\documentclass{sig-alternate}
\pdfoutput=1

 \usepackage{balance}
\usepackage{epsfig}
\usepackage{graphicx}
\usepackage{subfigure}
 
\usepackage{algorithm}
\usepackage[noend]{algorithmic}
\usepackage{multirow}
\usepackage{xspace}
\usepackage{url}
\usepackage{color}

\begin{document}
\title{On Integrating Information Visualization Techniques into Data Mining: A Review
}

\author{
\begin{tabular}{c}
Keqian Li
\end{tabular}\\
\begin{tabular}{c}
 {\sf keqianli@cs.ubc.ca }
\end{tabular}
}

\date{}
\maketitle

\begin{abstract}
The exploding growth of digital data in the information era and its immensurable potential value has called for different types of data-driven techniques to exploit its value for further applications. 
Information visualization and data mining are two research field with such goal. While the two communities advocates different approaches of problem solving,
the vision of combining the sophisticated algorithmic techniques from data mining as well as the intuitivity and interactivity of information visualization is tempting.
In this paper, we attempt to survery recent researches and real world systems integrating the wisdom in two fields towards more effective and efficient data analytics.
More specifically, we study the intersection from a data mining point of view, explore how information vis can be used to complement and improve different stages of data mining through established theories for optimized visual presentation as well as practical toolsets for rapid development. We organize the survey by identifying three main stages of typical process of data mining, the preliminery analysis of data, the model construction, as well as the model evaluation, and study how each stage can benefit from information visualization. 
\end{abstract}


\section{Introduction}
The exploding growth of digital datasets in the information era and its immensurable potential value has called for different types of techniques to extract its value for further data-driven applications. Most analytical techniques such as is commonly defined as neural networks, association rule, numerous clustering and classification methods are covered in the field of Data Mining, "the extraction of implicit, previously unknown, and potentially useful information from data."\cite{witten2005data} On the other hand, the field of Information Visualization(InfoVis) focuses on visual reproduction of the data. The rationale is that "visual representations and interaction techniques take advantage of the human eye's broad bandwidth pathway into the mind to allow users to see, explore, and understand large amounts of information at once." \cite{thomas2005illuminating}

Both fields hold the aim of processing real world data to facilitate further application. 
However there is a sharp difference between the two subject. InfoVis primarily focus on producing output of graphic format based on original structure of data which usually involves human sense making\cite{pirolli2005sensemaking} to further the analysis. Data mining, on the other hand, use automatic algorithmic techniques in such analysis process, to discover derived structure of data or directly provide off the shelf prediction for the objective of interest\cite{witten2005data}.

While the two communities advocates different approaches of problem solving and partly represent contrary side of the long term phylosophical debate of Hypothesis testing vs. exploratory data analysis\cite{herschel1831preliminary},  the noticecable overlap of ultimate goal as well as application scenario has lead to possible light of collaboration. 
The vision of combining the sophiscated algorithmic techniques from data mining as well as the intuitivity and interactivity of information vis is tempting. 
 
In this paper, we attempt to survery recent researches and real world systems integrating the wisdom in two fields towards more effective and efficient data analytics.
More specifically, we study the intersection from a data mining point of view, explore how information vis can be used to complement and improve different stages of data mining through established theories for optimized visual presentation as well as practical toolsets for rapid development. Previous

We organize the survey by identifying three main stages of typical process of data mining and study how each stage can benefit from information visualization. 
Data mining research usually come with the stage of \textbf{preliminery analysis of data} to show the basic characterstic of data before heavy data mining 
algorithm is applied, followed by the stage of \textbf{model construction}, then finally the \textbf{model  evaluation}. We devote section 2 to investigate how InfoVis can be of use in this context. More specifically, we explore in the task usually termed as \textbf{preliminery analysis of data}, how InfoVis can help human discover patterns by browse and navigation through the raw data. Then we devote section 3 to survey how InfoVis can help \textbf{interactive model construction} in which visualization is provided 
for partial results of the current construction of model while user can provide feedback to change the behavior of the process. We investigate \textbf{visualization of model evaluation} in section 4 to study how to use the InfoVis to convey the result intuitively to facilitate the process of understanding and further adjusting the model, and finally conclude in section 6.

\section{Related Work}
There exist several previous survey that have overlaps with our work includes \cite{}. Keim's survey \cite{} on Visual Analytics focuses on the different graphical visualization technique and how they are used in real visualization systems. Oliveira's survey \cite{} on Visual Data Mining gives an extensive discussion on basic terminologies in data analysis and data mining. Different visualization technique as well as Formal Models of Visualization are discussed. They also talked about maching learning model based visualization, where several example system are introduced. 
In contrast to above, our work take a data mining point of view, surveyed different how different information vis can be used to complement and improve the preliminery analysis of data, model construction, model  evaluation by providing comprehensive review with well organized taxonomy. 


\section{Visualization for Preliminary Data Analysis}
The task of data mining typically come with the a light weight stage of \textbf{preliminery analysis of data} before main data mining algorithm is applied. 
 
The goal is to show the basic characterstic of data and help form a quick grasp of intuitive understanding of the data to gather information. The insight formed in this stage can greatly help data miners to formulate the strategy of further heavy weighted analysis inspire and facilitate further data mining approach. 
The visualization preliminery analysis of data, involves techniques for visual represention, arrangement and simple manipulation the data. 

Here we provide an overview for the off-the-shelf visualization techniques both from recent data mining and visual data analytics research or classical statistical plot technique, grouped by different type of data to be visualized.

\subsection{Scalar value data visualization}

In this section we review some basic visualization techniques for visualizing single numbers, which serve the foundation for more complex patterns.
Here the important questions include how big the value is, or how big compared to other single values, or how the groups of values are distributed, or how it is changing according to external dimension such as the time. 
Here we allow the scalar value to be index by one or more attribues as the time in time-series data \cite{hamilton1994time}.
The visualization technique can be classified according to these different purposes.  

For a small number of values, graphically rendering may help to intepret how each individual value means by showing how big the values are and performing comparison across multiple values.
To serve these purpose, the technique of bar chart\cite{munzner2014vis} display values using horizontal aligned bars with length corresponding to the scale of value. It is is good for letting users focus on distinct scores indexed by a distinctive feature, such as nomial attributes.
To further group the data, Stacked bar charts to aggregate each group of values into one bar, 
using the length of bar as well as subcomponent to show the aggregation in addition to lookup individual values.

\subsubsection{Scalar value distribution}
For a large group of values, one may be interested in the overall trend and aggregation of these values instead of observing individual ones. More specifically, here we are interested in visualizing the distribution of scalar values.

A straightforward way is to plot the corresponding probability against different range of values. When the range is discretly partitioned into bins, the plot with each bar denoting different frequencies are called  histograms \cite{munzner2014vis}, alternatively we can render the graph of probability distribution function(p.d.f.) using standard function plot \cite{cleveland1985elements}.

\begin{figure}[t!h!]
  \centering
    \includegraphics[width=.4\textwidth]{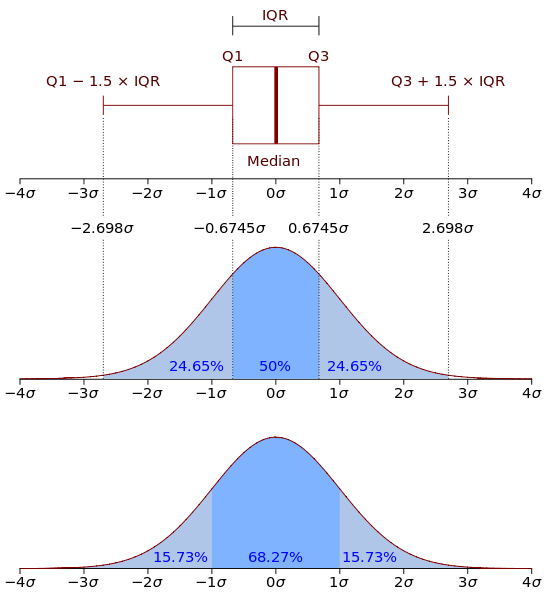} 
	\caption{showing the same distribution with boxplot and probability density function}
\label{fig:localrank}
\end{figure}

The the box and whiskers plot is concise idiom of plots for showing distribution using a box with emphasis on some important quantitive attribute including the median, the first and third quartiles and the chosen threshold for outliers \cite{mcgill1978variations}. Outliers are shown as individual dots. The boxplot is good for showing 
the attributes of a distribution such as spread, skewness, as well as comparing across multiple distribution.

There are many variants of boxplots that
augment the basic visual encoding with more information. 
Variable width box plots illustrate the size of each group whose data is being plotted by making the width of the box proportional to the size of the group. A popular convention is to make the box width proportional to the square root of the size of the group. Important variants using different density esitimation methods and preprocessing include the vase plot and voilin plot. \cite{mcgill1978variations}
A bagplot, or starburst plot \cite{rousseeuw1999bagplot} is a method in robust statistics for visualizing two-dimensional statistical data, analogous to the one-dimensional box plot. It uses shapes of different enclosing polygons to visualize the location, spread, skewness, and outliers for a two dimensional distribution.
 
In statistical graphics, the functional boxplot is an informative exploratory tool that has been proposed for visualizing functional data.\cite{sun2011functional} Again analogous to the classical boxplot, it uses of the envelope of the 50\ 

\subsubsection{visualizing scalar value along other dimensions}
Scalar values usually come along with other attributes of the data, one particular interest is to study the relation between the scalar value and other dimensions of data, for example time series analysis for stock price. To serve the purpose of how the scalar value is changing according to the external dimensions, we describe the family of scatterplots for one dimension index as well as techniques for two dimension index or spatial index.

\begin{figure}[t!h!]
  \centering
    \includegraphics[width=.4\textwidth]{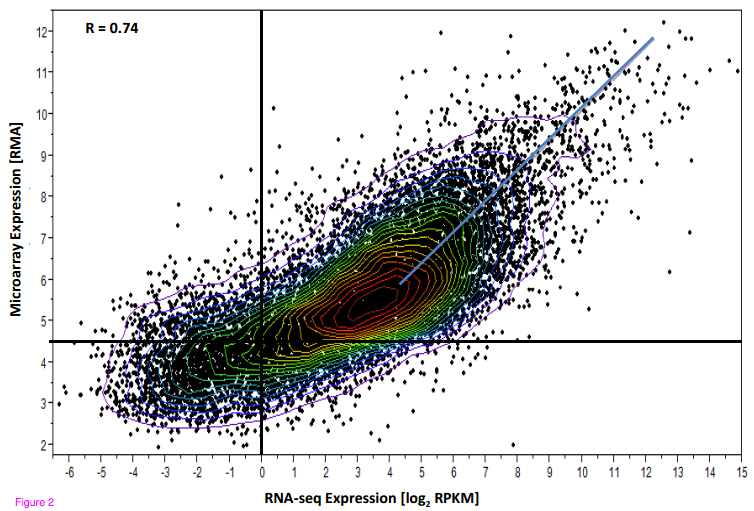} 
	\caption{a 2d scatterplot with contour and a calculated regression line}
\label{fig:localrank}
\end{figure}

The idiom of scatterplots \cite{jarrell1994basic} encodes two index dimension using
both the vertical and horizontal spatial position, and the each data is represented with a point mark.
Scatterplots are effective for the abstract tasks of providing overviews
and characterizing distributions, and specifically for finding outliers and
extreme values. Scatterplots are also highly effective for the abstract task
of judging the correlation between two attributes.
On the scatterplots, the derived
data of a calculated regression line is often superimposed on the raw scatterplot
of points.
Additional transformations such as log-plot are useful for better comparing the relationship.
More generally, the idiom of banking to 45 degree computes the best aspect ratio, i.e. the the ratio of width to height, for a chart
in order to maximize the number of line segments that fall close
to the diagonal.
When the external dimension serves as index for the scalar value, it is also called dot chart\cite{munzner2014vis}. If we further incorporate 
line connection marks running between the points to emphasize the trend through the dimension of index, it's called line charts. One important difference between these two is that dot chart is more applied to nominal index while  line charts is for ordered index.

For the case that two dimension index, we encode the each index dimension as different axes in a matrix, with the value encoded using color channel. An important component is the how to map values into color. There are many different such techniques color schemes that can be used to illustrate the heatmap, with perceptual advantages and disadvantages for each \cite{harrower2003colorbrewer, green2011colour}.
Another special case is the spatial data index, where the external dimensions map to specific 2 dimensional or 3 dimensional geographic locations. For spatial data, the external dimension naturally lend itself to the coordinates for rendering. For the value itself, one way of handling is to map it to a non-spatial visual channel, e.g. the color in Choropleth Maps. Alternative way to introducing a new channel but to visualize the level set \cite{stewart2011multivariable} for each specific level of the value along the dimensions, such as topographic terrain map in geometry in for the case of 2 dimensional index \cite{davis2002statistics}. In 3 dimensional, however, occlusion becomes a major problem. Interactive method is commonly introduced to let user observe different level set at different time, or we can carefully choose a subset of level sets with little occlusion to be shown at once \cite{kniss2001interactive}.

\subsection{Multi-dimensional data visualization}  
 
The analysis of Multi-dimensional data requires inspection of multiple attributes at the same time to investigating joint semantics meaning, or specific types of relations between different dimensions. 
One natural idea is to treat the problem as visualizing vector value indexed by different other dimension as opposed to visualizing scalar value, use the similar encodings techniques of coordinate but the specific types of encoding for vectors, of which the major technique is glyph. Alternatively, we take a wholistic view of all the columns of values to be shown and 
visual each columns with heterogenous encoding to facilitate comparison and cognition across every column of the value. In addition to the general purpose techniques above, we also investigate  visualization technique for the specific relation of bipartite flow.

\subsubsection{glyph based vector visualization}
The glyph is a general term for visualization of a multidimensional data record which can be placed in a specific location in the plot. \cite{ward2008multivariate} It has the ability to encode a set of values in contrast to a simple point mark that only encodes the location and therefore is a general way of augmenting typical point mark based visualization. That is, instead of visualizing a scalar field, here we are visualizing a vector field or even a tensor field. A glyph can be designed by associate the element of vector with 
position (1, 2, or 3-D), size (length,
area, or volume), shape, orientation, material (hue, saturation, intensity, texture,
or opacity), line style (width, dashes, or tapers), and dynamics (speed
of motion, direction of motion, rate of flashing).\cite{ward2008multivariate, borgo2013glyph}

\begin{figure*}[t!h!]
  \centering
    \includegraphics[width=.9\textwidth]{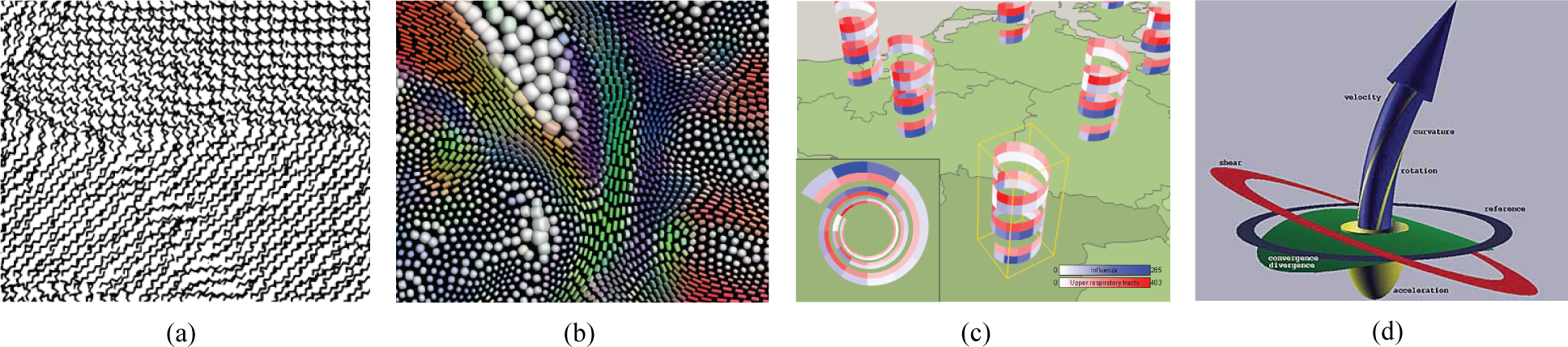} 
	\caption{Several example of glyphs (a) Stick figures form textural patterns [PG88]. (b) Dense
glyph packing for diffusion tensor data [KW06]. (c) Helix glyphs on maps for analyzing cyclic temporal patterns for two
diseases [TSWS05]. (d) The local flow probe can simultaneously depict a multitude of different variables [dLvW93]}
\label{fig:localrank}
\end{figure*}

The using of glyph is ubiquitous. In the scatterplot, if we use glyph at Each point to incorporate additional visual channels such as the size, we will have bubble chart\cite{munzner2014vis}. 
 
For the task of showing 2 dimension vector field, we can render each vector using the technique of flow glyph\cite{wittenbrink1996glyphs}, which usually encodes magnitude with the length of the stem, direction with arrow orientation, or more complex patterns such as radial axes \cite{hlawatsch2011flow}.
More generally, we can also visualize a tensor field by encoding a matrix as a glyph\cite{kindlmann2004superquadric}.
Tensor fields typically contain a matrix at each cell in the field, capturing
more complex structure  such as stress, conductivity,
curvature, and diffusivity.
To augment the vector we can use a geometric
shape to encode the 
the eigenvalues of the tensor matrix
from the eigenvectors
with shape, orientation, and further visual channels.
Glyph can also be incorporated with icons for better expressiveness. The Chernoff face \cite{chernoff1973use} is a famous example for encoding multivariate attribute as a cartoon face, where different parts of the face including eyes, ears and nose are used to encode values of the variables by their shape, size and orientation.

\subsubsection{similarly encoding each dimension}
Instead of vector value index by different dimensions, we can treat the whole data as a multivariate table to 
visual each columns with heterogenous encoding to assist comparison and cognition across every column of the value.

A scatterplot matrix (SPLOM) is a matrix of scatterplot showing pairwise relation for the attributes.
The key is a simple derived attribute that is the same for both the rows
and the columns: an index listing all the attributes in the original dataset.
SPLOMs are heavily used for the abstract tasks of finding correlations,
trends, and outliers, in keeping with the usage of their constituent scatterplot
components. Certain
systems, such as \cite{elmqvist2008rolling}, allow the rows and columns of a SPLOM
to be reordered, for added flexibility
.

In addition to each off-diagonal plot mapping a pair of non-identical dimensions, \cite{cui2006enhancing} study the use of diagonal plots. In their work, histograms, 1D plots and 2D plots are drawn in the diagonal plots of the 
scatterplots matrix. In 1D plots, the data are assumed to have order, and they are projected in this order. In
2D plots, the data are assumed to have spatial information, and they are projected onto locations based on these
spatial attributes using color to represent the dimension value. The plots and the scatterplots are linked together
by brushing. Brushing is applied on the diagonal plots and regular scatterplots, 
together with other further visualizations in the system, including parallel coordinates and glyphs.

In \cite{viau2010flowvizmenu} they develop several additive elements of the 
a scatterplot matrix (SPLOM). 
Their first variant of the SPLOM can order scatterplots according to some
ranking or metric of interest.
Secondly, they described a novel arrangement of scatterplots
called the Scatterplot Staircase (SPLOS) that requires less space than a traditional scatterplot matrix.
It works by showing only the scatterplots of consecutive pairs of
dimensions, arranged in a staircase pattern, such that adjacent scatterplots
share an axis along their common edge.
Lastly, they provide a hybrid technique for a scatterplot matrix (SPLOM)
and parallel coordinates called the Parallel Scatterplot Matrix (PSPLOM)
based on the idea that rotation around the vertical axes causes the visualization to transition
from a sequence of scatterplots to a PCP.

Another major family of high dimensional
data visualization is the Parallel Coordinates techniques, in particular for data mining purposes 
\cite{achtert2013interactive}.
Parallel Coordinates encode dimensions as vertical axes Parallel to each other, rather than perpendicularly at
right angles. 
For each data instance is encodes as a polygonal line which intersects vertical axes at 
corresponding locations, rather than a point at specific horizontal and vertical positions.
Parallel coordinates are more often used for other
tasks, including overview over all attributes, finding the range of individual
attributes, selecting a range of items, and outlier detection.
For example, one can observe the correlation across multiple attributes by checking the directions and angles of different polygonal lines of the data instances.
In Parallel Coordinates, the order of the axes is critical for finding features, and in typical data analysis many reorderings will need to be tried. Some authors have come up with ordering heuristics which may create illuminating orderings.\cite{wang2003interactive}.

Instead of a polygonal line, a smooth parallel coordinate plot is encode each data item with splines \cite{moustafa2002some}. In the smooth plot, every observation is mapped into a parametric line (or curve), which is smooth, continuous on the axes, and orthogonal to each parallel axis using shape manipulation. This design emphasizes the quantization level for each data attribute \cite{moustafa2006multivariate}.

To conquer the major challenge is the ordering of axes in parallel coordinate plot, as any axis
can have at most two neighbors when placed in parallel on
a 2D plane, \cite{achtert2013interactive} extend this concept to a 3D visualization
space so they can place several axes next to each other. 
They provide a system to explore
complex data sets using 3D-parallel-coordinate-trees, along
with a number of approaches to arrange the axes.

\begin{figure}[t!h!]
  \centering
    \includegraphics[width=.4\textwidth]{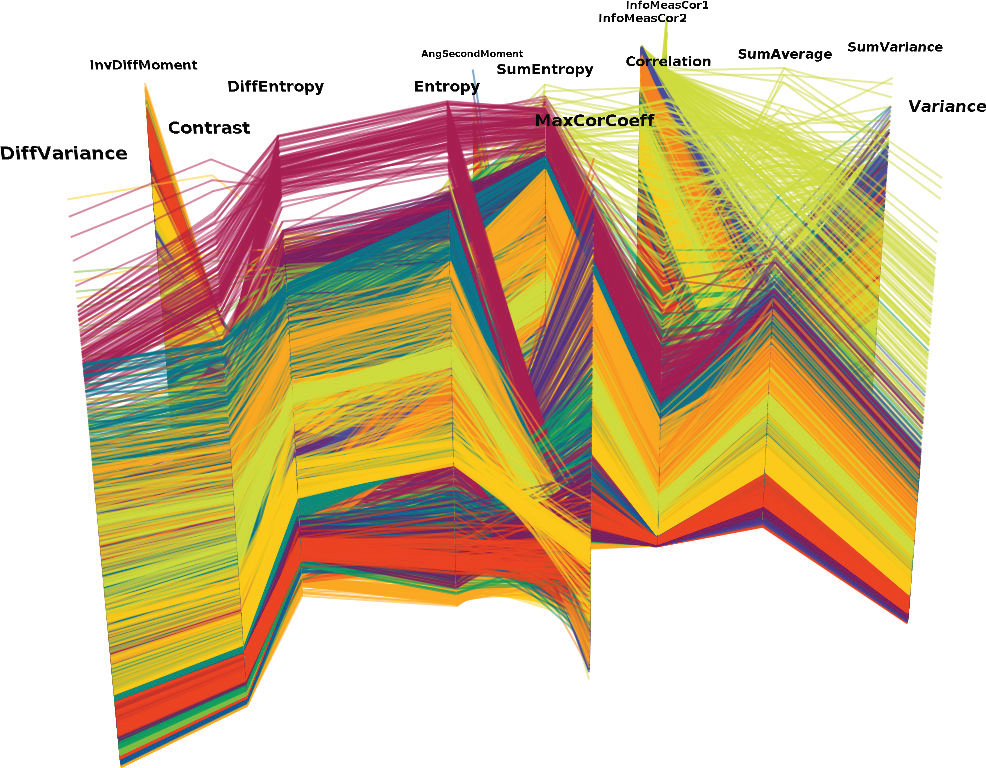} 
	\caption{illustration of 3D-Parallel-Coordinate-Trees}
\label{fig:localrank}
\end{figure}

\begin{figure}[t!h!]
  \centering
    \includegraphics[width=.4\textwidth]{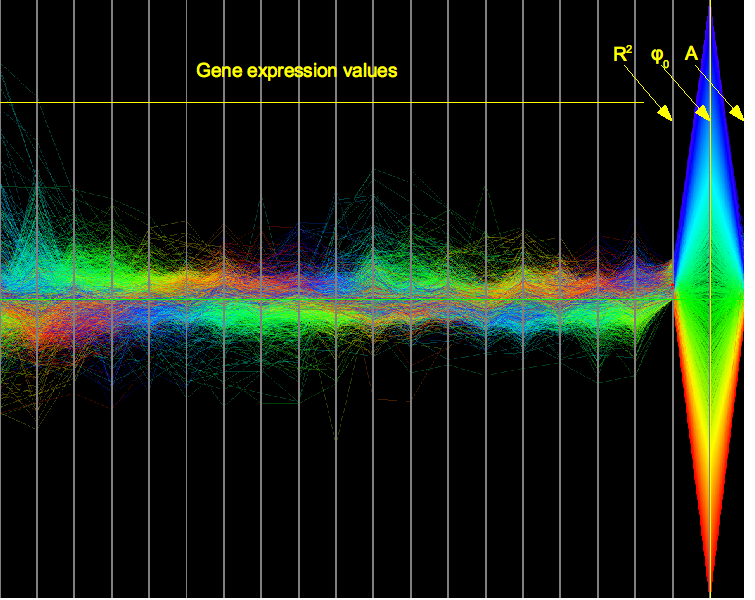} 
	\caption{SpRay analysis on Yeast cell cycle dataset}
\label{fig:SpRay}
\end{figure}

As an application on visual exploration of
large and high-dimensional real world datasets, the authors of \cite{dietzsch2009spray} applied parallel coordinate plot based Visual Analytics on  multidimensional gene expression datasets. Their approach is to visualize the gene expression conjoined with statistical data together in the parallel coordinate plot.
They present a new application, SpRay, designed for the visual exploration
of gene expression data. They investigate
the visual analysis of gene expression data as generated by microarray
experiments, combine refined visual exploration with statistical
methods to a visual analytics approach that proved to be
particularly successful in this application domain. 
Figure \ref{fig:SpRay} The first 18 dimensions of this parallel coordinate plot (PCP) correspond to the gene expression values
at the 18 time points for the a factor arrested cells. The last three dimensions correspond to the values
obtained by the harmonic regression analysis (HRA): the coefficient of determination, the zero-phase angle, and the amplitude of the
estimated curve. The polylines of the PCP are colored by the zero-phase dimension, so that the periodic changes of
the transcript levels of groups of genes can be very easily identified.
They demonstrate
the usefulness on several multidimensional gene expression
datasets from different bioinformatics applications such as finding periodic patterns in
microarray data, emphasizing relevant
expression patterns, outlier detection.

\subsubsection{3D Visual Data Mining}

\begin{figure}[h]
\centering

\subfigure[Scatterplots]{
   \includegraphics[width=.2\textwidth] {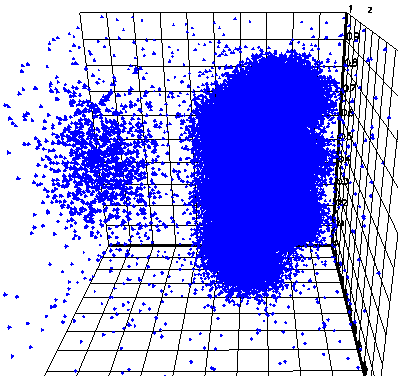}
   \label{fig:subfig1}
 }
 \subfigure[Equalized Density Surfaces]{
   \includegraphics[width=.2\textwidth] {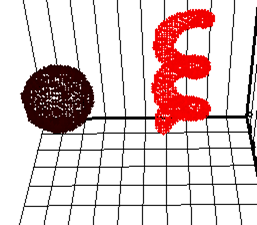}
   \label{fig:subfig2}
 }

\label{fig:3dvdm}
\caption{Equalized Density Surfaces technique for 3DVDM}
\end{figure}
The 3D Visual Data Mining (3DVDM) is a promising research area that map data records into a 3D place and use geometric method for analysis.

 \cite{mazeika2008using} introduce and evaluate nested surfaces for the purpose of
3D visual data mining. 
Nested surfaces enclose the data at various density levels and facilitates the detection of multiple structures, which is important for data
mining where the less obvious relationships are often the most interesting ones.
They give
a topology-based definition of nested surfaces and establish a relationship to the
density of the data. Algorithms are given that compute surface grids and
surface contours based on an estimated PDF which makes our method independent
of the data. Real time
interaction can be achieved by precomputing and storing small density estimates.

\cite{bohlen20083dvdm} investigated the different 3D Visual Data Mining technique and illustrate the use with clickstream data analysis. They try to leap the step from statistical
charts to comprehensive information about customer behavior by proposing a density surface based analysis of 3D data that uses state-of-the-art interaction techniques for interpretation of the data at the conceptual level. 
Animation, conditional analyzes, equalization, and
windowing are crucial interaction techniques in their system, making it possible to explore the data
at different granularity levels, which leads to a robust interpretation.
Fig. \ref{fig:3dvdm} illustrates the use of equalize the structures in support of the exploration
of not equally pronounced structures, which clearly shows that the Equalization of Structures helps to identify structures in the data.

\subsection{Bipartite flow visualization}
In this section we investigate the visualization technique for a specific type of relation between datas, the flow intensity for two bipartite dimensions, to show how visual idiom can be developed based on the nature of specific type of relations.

\begin{figure}[t!h!]
  \centering
\subfigure[Computed Lattice for a bipartite relationship]{
   \includegraphics[width=.4\textwidth] {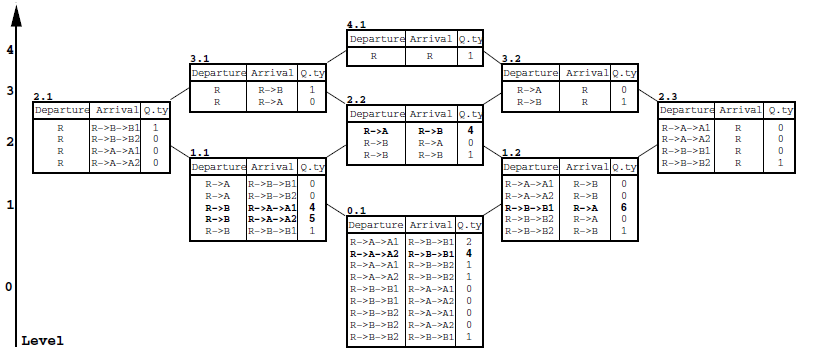}
   \label{fig:subfig1}
 }
 \subfigure[3D visualization for binary heavy hitters]{
   \includegraphics[width=.4\textwidth] {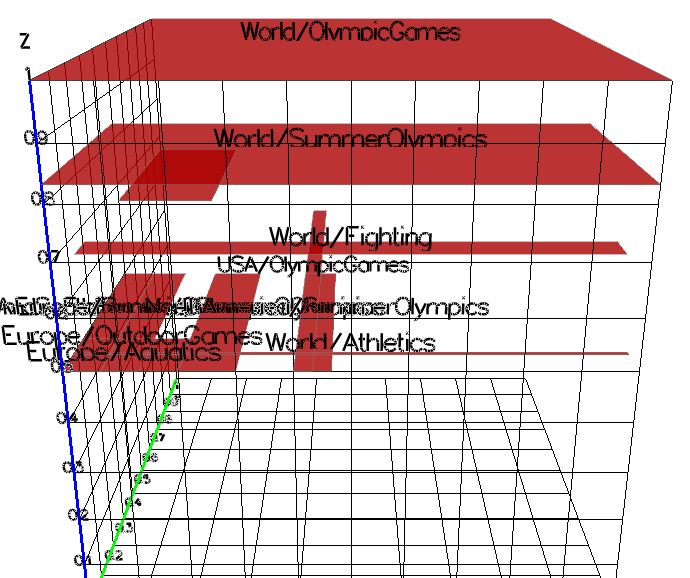}
   \label{fig:subfig2}
 }
	\caption{Hierarchical heavy hitters visualization}
\label{fig:localrank}
\end{figure}

\begin{figure}[t!h!]
  \centering
    \includegraphics[width=.4\textwidth]{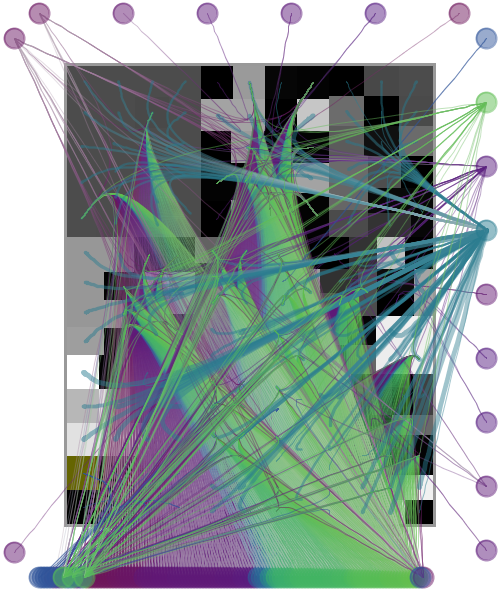} 
	\caption{Nflowvis for network traffic analysis}
\label{fig:nflowvis}
\end{figure}
Hierarchical heavy hitters on for two-dimensional data is a widely adopted technique for identification of significant relationship between a large set of bipartite relationships, developed from the concept of hierarchical heavy hitters (HHHs) from database theory \cite{cormode2003finding}. 
\cite{mazeika2008analysis} provide a visualization tool to compute and investigate hierarchical heavy hitters for two-dimensional data. 
 Given a two-dimensional categorical dataset with hierarchical
attribute values and a threshold, the task is to visualize the HHHs in the three dimensional space. 
They first perform aggregation to reduce the large set of binary relationships into a lattice through grouping, filtering and balance, and encode each hierarchy of attributes with different Level Planes 
for each of the two part of the entities where conceptual inclusion in indicated by one covering another in 3D spaces.

\cite{mansmann2009visual} developed a visual analytics system of NFlowVis to investigate the network flow between different local hosts and external IPs, with application in real world network attack detect tion in computer network.
They introduced the home centric flow visualization where the local hosts that are related to attacking hosts are visualised in a
TreeMap and attacker hosts are placed at the borders. Flows between
attackers and local hosts are visualised using splines. The colour of the local
hosts and their size and the colour of the splines can be used to represent various
properties, such as packets or bytes transferred. Thresholds can be used
to hide splines with a low traffic and highlight splines with a high traffic to the
attackers as shown in the screenshot of NFlowVis in Figure \ref{fig:nflowvis}.
They also proposed graph-based flow visualisation. 
The main advantage of the graph view is that it emphasizes
structural properties of the connectivity between hosts, such
as groups of interconnected hosts.

\subsection{Network Data}
Network Data arises in many important applications, such as from online social media, and physical or biological interaction network, and more generally a set of entities with interdependencies among others , underscored by the surge of social media over the recent years, such as Twitter and Facebook \cite{cohen2010complex}.
Due to its sheer importance, tons of work has been done for visualizating such kind of data. There are many widely used graph visualizing systems including UCINet, JUNG, and GUESS, GraphViz, Pajek, Visone \cite{borgatti2002ucinet, o2005analysis, huisman2005software}, used by different communities.
Two main family of Network Data encoding techniques are node-link diagrams and adjacency matrix views. 

\subsubsection{Node-link diagrams}

\begin{figure}[t!h!]
  \centering
    \includegraphics[width=.4\textwidth]{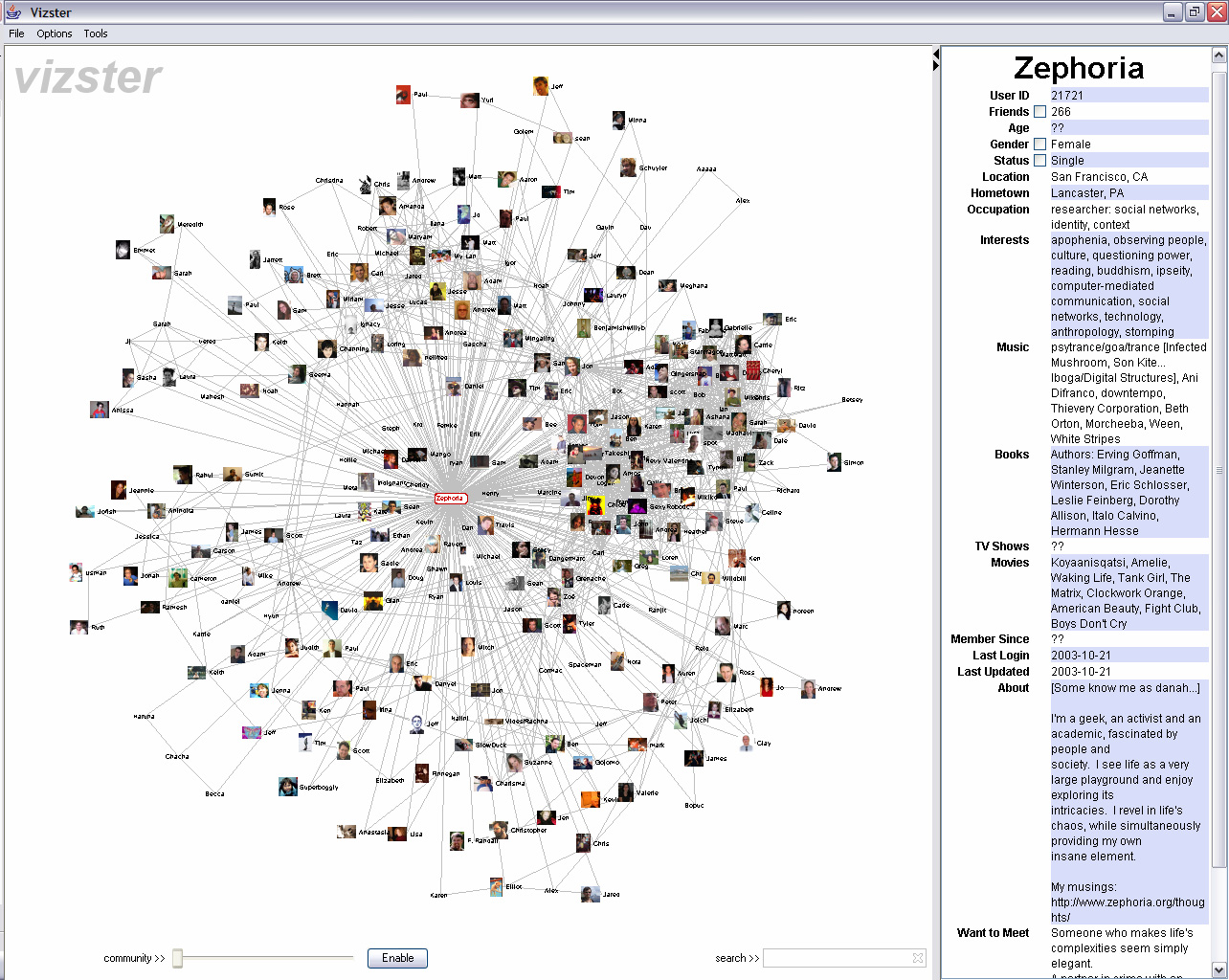} 
	\caption{Vizster for attribute search enhanced node--link diagrams}
\label{fig:Vizster}
\end{figure}

A straightforward way of encoding nodes and edges in network data
is with node–link diagrams, where nodes are drawn as point marks
and the links connecting them are drawn as line marks.
One important choice of visualization is the how to arrange the layout of nodes and links in the plot, which is crucial for revealing connection patterns such as densely connected communities structure in the network. Common layout techniques for networks include simple grid method such as Orthogonal layout, Arc diagrams, Circular layout
graph structure based method such as Layered graph drawing, Spectral layout methods and Dominance drawing,
physical computation driven methods such as force based layout and Annealed layout\cite{brandes2001drawing, davidson1996drawing, von2011visual, tamassia2011handbook}. 
Since many visualization task need to deal with very large network, how to effectively present such amount of data becomes an essential challenge.
reducing technique becomes necessary. It is typically done by aggregating nodes into communities, and treat these as supernodes to form a reduced graph upon. Multilevel hierarchy can be incorporated through techniques such as FM\textsuperscript{3}, TopoLayout, sfdp \cite{hu2005efficient, hachul2005drawing, archambault2007topolayout}.

There are a large number of network visualization systems incorporating the node-link diagram together with further visualization module for nodes and links incorporating richer meta attributes. One good example is Vizster \cite{heer2005vizster}. They have designed a complex view builds upon familiar node-link network layouts to contribute customized techniques. Important techniques include exploring connectivity in large graph structures with graded color scale encoding on the path, supporting visual search and analysis with keyword search and attribute filtering, and automatically identifying and visualizing community structures.
Fig. \ref{fig:Vizster} shows a screen shot of the main interface Vizster visualization system. The left side is the main network display with controls for community analysis and keyword search. The right side consists of a profile panel showing a selected node's meta information. Words in the profile panel that occur in more than one profile will highlight on mouse-over; clicking these words will initiate searches for those terms. The checkboxes in the profile panel will initiate an "X-ray"
 view of that particular profile dimension.

Special effort has been on using different link view technique on real world social network exploration for specific analytic purpose.
\cite{simoff2008visual}  demonstrated the reflection of the analytical process in real world case study for social network analysis on link view
More specifically, the case illustrates the use of visual discovery for identifying fraudulent activity in
terms of the profile of the network patterns between attributes in the data set. The input of the 
brief from the company was essentially open-ended and without specification, making the case in the realm of ‘discovery' as there is little or no preliminary knowledge of
what are fraudulent patterns. They use the link view technique of overview, close-up, graphical irregularity discovery, attribute visualization, detail information panel, link navigation to illustrate the discovery process of 
Visual Models Generation, Cognition and sense-making and Discovery.
Similar link view analytics is done by \cite{huang2009visualization}. They visualize the network trading pattern in stock market, analyze suspected behavior of domination of
the stock price on the network, and pull out historical graph to identify frauds by patterns matching.

\subsubsection{Adjacency matrix view}

Social networks usually display a locally
dense pattern, which makes node-link displays less effective on these regions. 
An alternate way of visualizing network is or adjacency matrix view, where the connection pattern is 
treated as an adjacency matrix. The visualization then arrange nodes as rows and columns of a matrix and visualize 
different cell of the matrix according to the corresponding adjacency matrix element.
One important design choice in the visualization is the the ordering of such columns, which is usually achieved through ranking by attributes, ranking by cluster, by the sequence of a traversal algorithm such as TSP \cite{henry2006matrixexplorer}.

\begin{figure}[t!h!]
  \centering
    \includegraphics[width=.4\textwidth]{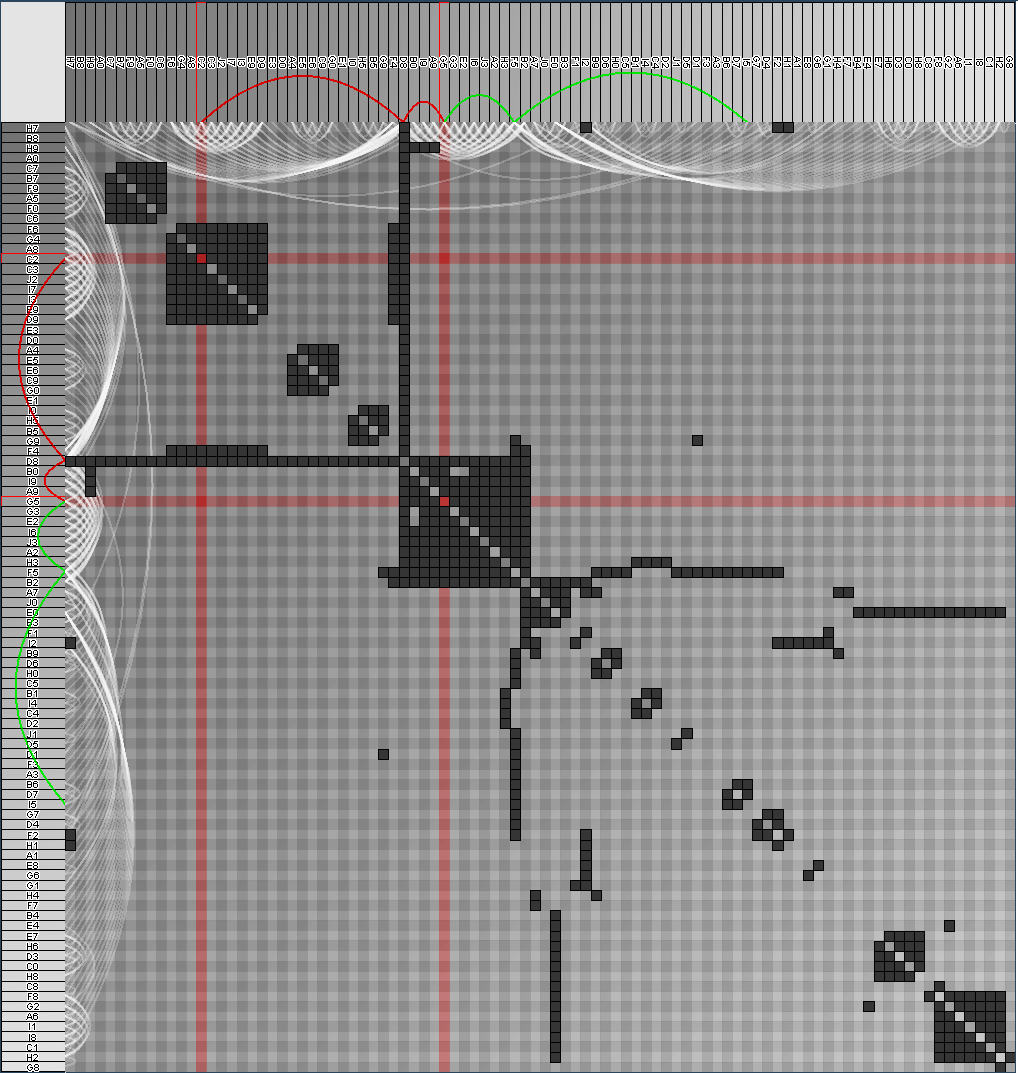} 
	\caption{MatLink for links and highlight enhanced matrix view}
\label{fig:localrank}
\end{figure}

\begin{figure}[t!h!]
  \centering
    \includegraphics[width=.4\textwidth]{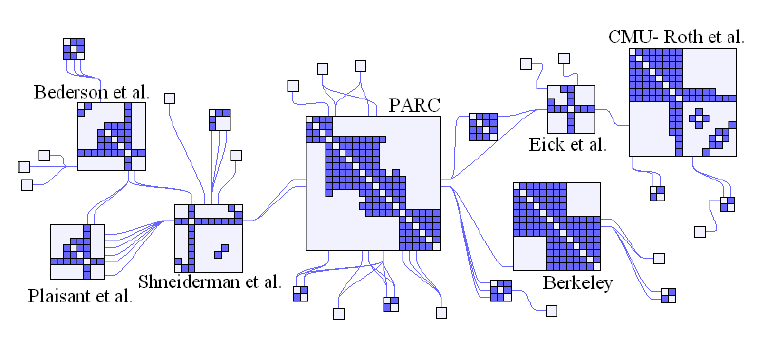} 
	\caption{NodeTrix using matrix view for local communities analysis}
\label{fig:NodeTrix}
\end{figure}

\cite{henry2007matlink} presents a comprehensive adjacency matrix view based visualization system incorporating several further augmentation.
Their system MatLink , a matrix representation
with links overlaid on its borders and interactive drawing of additional
links and highlighting of the cells included in a path from the cell under the
mouse pointer.
Many common analysis
tasks require following paths throughout the network, which is difficult on matrices. 
They propose the module of link augmentation.
MatLink displays the full graph using a linearized node-link representation
called the full linear graph. Its links are curved lines drawn interior
to the vertex displays at the top and left edges of the matrix. Links are drawn
over the matrix cells, using transparency to avoid hiding them. Longer links
are drawn above shorter ones. The linear graph conveys detailed and long-range
structure together without hiding any detail of the matrix: a feeling for link
densities and sub-graphs, but also paths and cut points.
In addition they also introduce the heavy interaction to the matrix view.
When the user has selected a vertex in the rows or columns, it is highlighted
in red, and the shortest path between this vertex and the one currently under
the mouse pointer is drawn in green on the vertex area, mirror-imaged to the
links drawn in the matrix border.

Another interesting direction is the hybrid representation for networks that combines the advantages of both representations: node-link diagrams are used to show the global linkage pattern, adjacency matrices to better  reveal connection pattern in local communities.
The system of NodeTrix \cite{henry2007nodetrix} present such a technique to partially import the the matrix view of network
to resolve the basic dilemma of being readable both for the link view for global structure of the network and also for matrix view analysis of local communities. 
they also allow user to create a NodeTrix visualization by dragging selections to and from node-link and matrix forms. Fig. \ref{fig:NodeTrix} shows a case study on InfoVis 2004 coauthorship dataset to identity three types of different inner community structure, the Cross pattern between Shneiderman and his collaborators, Block pattern between Researchers at Berkeley, Intermediate pattern between Roth and his collaborators at
CMU.

\subsection{Text Data}
The booming of the World Wide Web comes with enormous amount of unstructured data beyond the multidimensional framework described above. One important type of such data is text.
Most text analysis method involved a vectorization approach with a "bag of words" assumption  which 
typically involve tokenization of the raw text, normalization or standardization of the
tokens, and finally selection of the final features from the results. 
 
However, the sequence of words and the natural context is also important for aiding such analysis.  
We organize the text analysis method into two clusters. First we study the visualization for text analysis techniques on high dimensional space. Next we focus on visualization for analysis techniques with dimension reduction.

\subsubsection{Analysis on high dimensional space}
A documenht can be thought of a sequence formed by elements from a huge vocabulary. In this section we disciss some text analysis techniques that keep the high space nature as is.

\begin{figure}[t!h!]
  \centering
    \includegraphics[width=.4\textwidth]{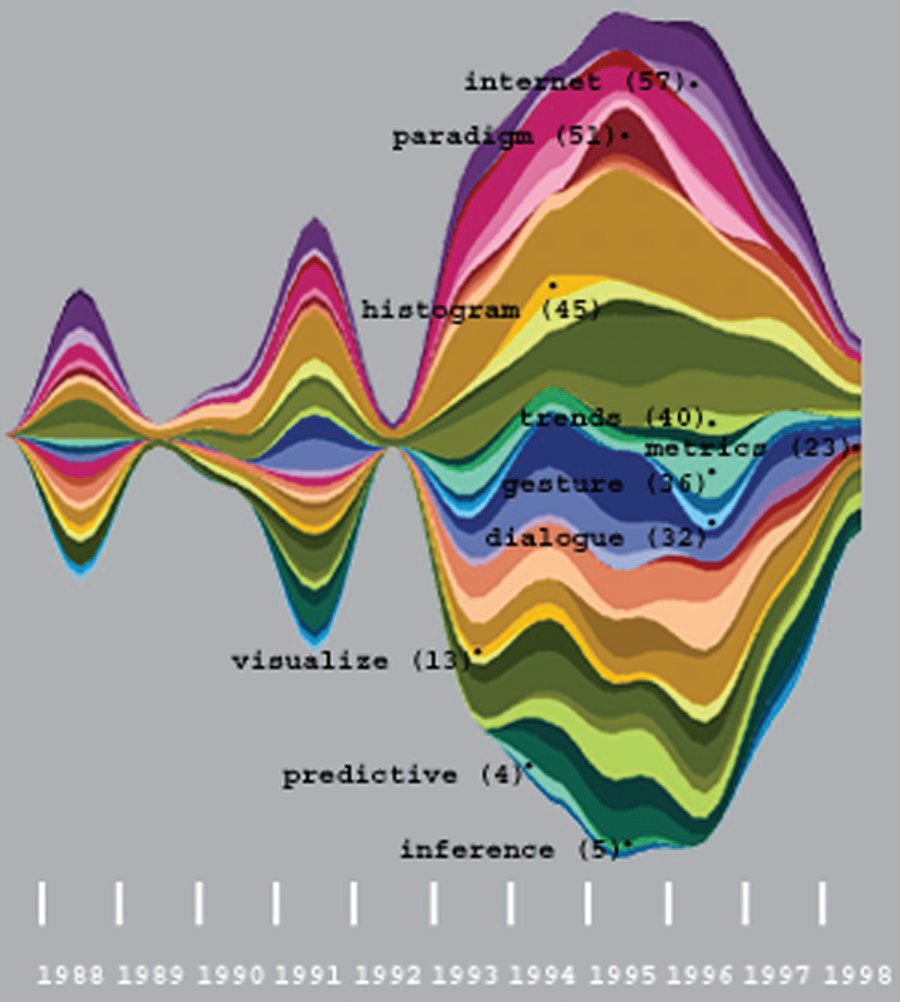} 
	\caption{ThemeRiver system using the word frequency vector to analyze the text}
\label{fig:localrank}
\end{figure}

The word frequency vector for a document is a straitforward and effective way of capture information from text, especially when the feature set is carefully chosen.
The ThemeRiver \cite{havre2002themeriver} system is an example of using the for text analysis. They captures the document using a frequency vector for a well chosen vocabulary, each indicating a specific "theme". They then visualize the changes in these "thematic" words in the context of a time line and corresponding external events to allows a user to discern patterns that suggest relationships or trends. They use a river metaphor to convey several key notions. The document collection's time line, selected thematic content,
and thematic strength are indicated by the river's directed flow, composition, and changing width, respectively. Colored
currents flowing within the river represent individual themes. A current's vertical width narrows or broadens to indicate decreases or
increases in the strength of the individual theme.

\begin{figure}[t!h!]
  \centering
    \includegraphics[width=.4\textwidth]{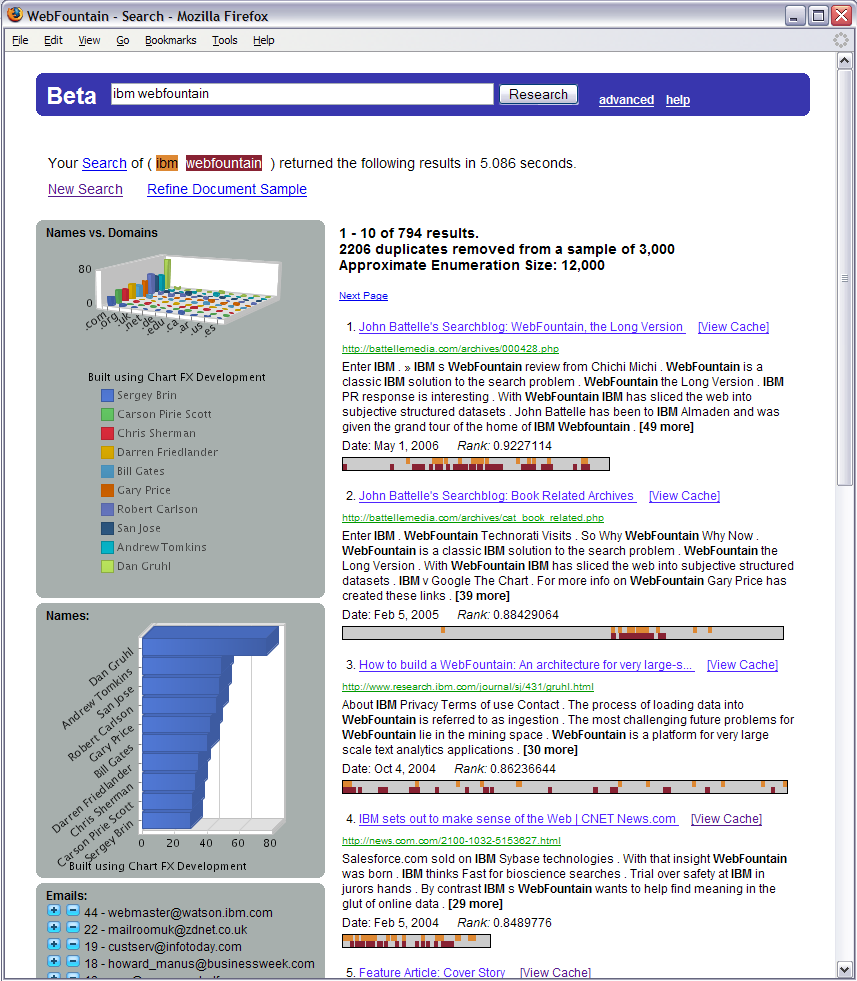} 
	\caption{The BETA system using entity based analysis to help the search query visualization}
\label{fig:BETA}
\end{figure}
Some visualization took advantage of semantic analysis technique.
 For example, the TAKMI text mining system\cite{nasukawa2001text}, in which text from call center complaints were analyzed to help staff members determine which problems with a product receive increasing numbers of complaints over time. The interface shows the distribution of entity mentions over time, using the brushing-and-linking technique to connect selected topics to bar charts.
The BETA system \cite{meredith2006beta} used entity based analysis to help the search query visualization. Figure \ref{fig:BETA} shows the results of a query on ibm webfountain. The search results list on the right hand side was augmented with a TileBars display \cite{hearst1995tilebars} showing the location of term hits in color and the relative lengths of the documents. Along the left hand side was shown a 3D bar chart of occurrences of entity names plotted against Web domains, as well as a sorted bar chart showing frequency of entity names across domains.

\begin{figure}[t!h!]
  \centering
    \includegraphics[width=.4\textwidth]{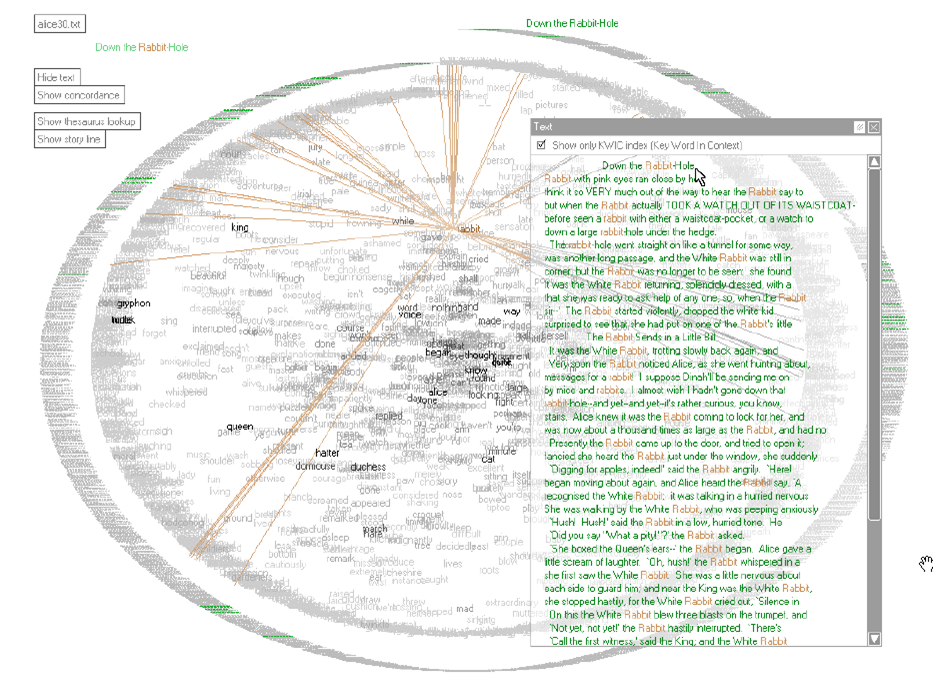} 
	\caption{TextArc system for concordance visualization}
\label{fig:TextArc}
\end{figure}
In the field of literature analysis it is commonplace to analyze a text or a collection of texts by extracting concordance: an alphabetical index of all the words in a text, showing those words in the contexts in which they appear.  The Word Tree \cite{wattenberg2008word} display the concordence relation using a tree structure, allowing the user to view words or phrases which precede or follow a given word, thus showing the contexts in which the words appear.
The TextArc  \cite{paley2002textarc} visualization arranged the lines of text in a spiral and placed frequently occurring words within the center of the spiral. Selecting one of the central words drew lines radiating out to connect to every line of text that contain that word, as shown in Fig. \ref{fig:TextArc}. Clicking on the word showed the contexts in which it occurred within the document, so it acts as a kind of visual concordance tool.

\subsubsection{Analysis with dimension reduction}
Many text analysis techniques treat the text as a word frequency vector and exploit numeric dimension reduction on the word vector for projection to low dimension space either to extract latent semantics or to prepare for 2D space display.

\begin{figure*}[t!h!]
  \centering
    \includegraphics[width=.8\textwidth]{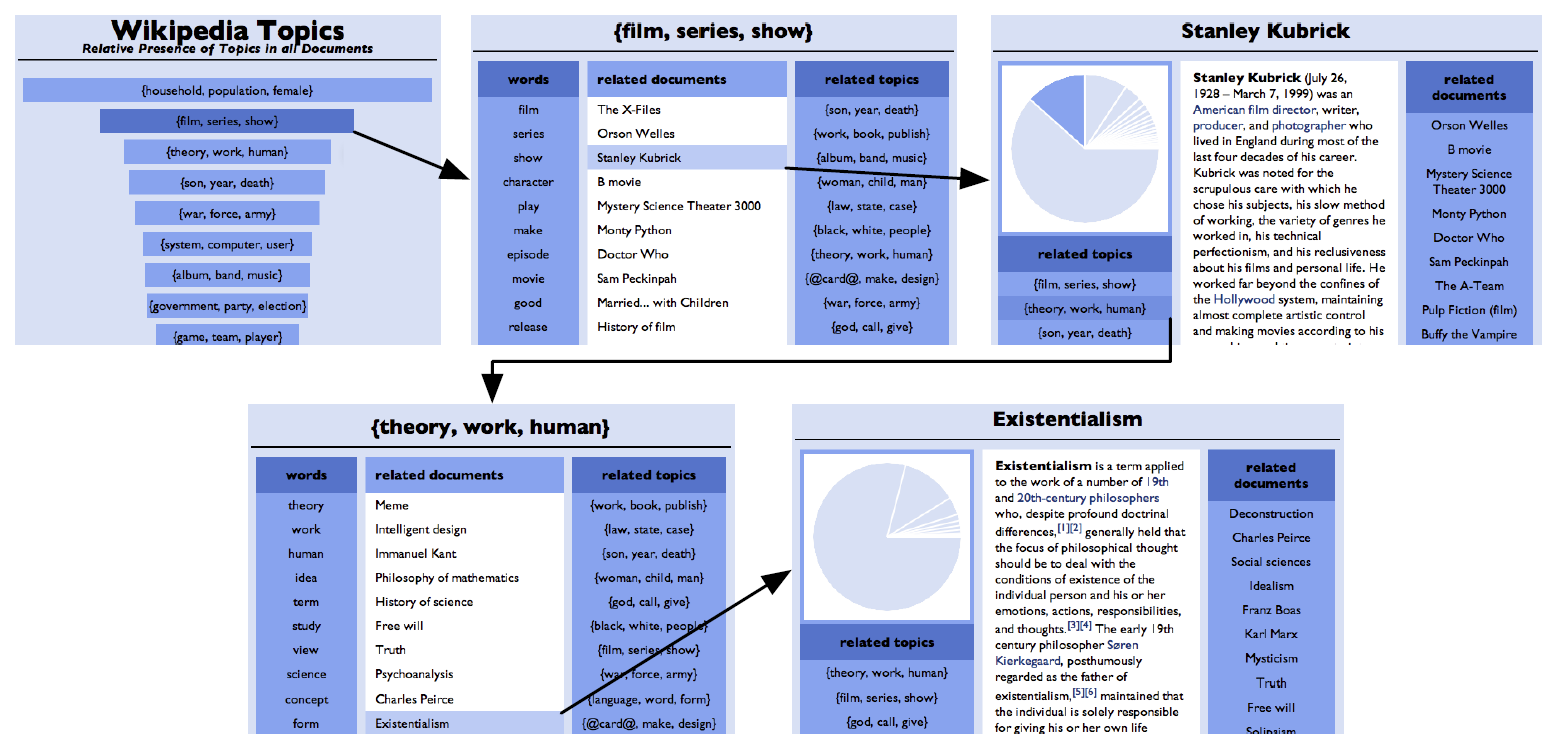} 
	\caption{Navigating flow based on the topic model fit to Wikipedia}
\label{fig:TM}
\end{figure*}

\begin{figure}[t!h!]
  \centering
    \includegraphics[width=.4\textwidth]{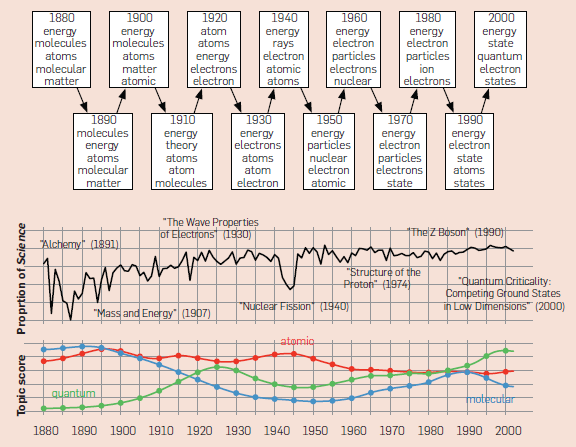} 
	\caption{one topics from a dynamic topic model fit to Science from 1880
to 2002}
\label{fig:dynamicTM}
\end{figure}

The LDA topic model \cite{blei2003latent} is a popular data mining method for analyzing documents in which the latent low dimensional space is interpreted as topic space. More concretely, the method learns a representation of document in terms of "topics", which in turn are defined as distribution over a fixed vocabulary. 
Sophisticated visualization based on interactive methods has been proposed. 
\cite{chaney2012visualizing} creates a navigator of the documents,
allowing users to explore the hidden structure exploration for the topic model 
by navigating through the dataset through the mapping of document to topics, topic to words as well as topic based similarities.
Fig. \ref{fig:TM} shows the flow of Navigating Wikipedia with a topic model. 
Beginning in the upper left, the users see a set of topics, each of which is a theme discovered
by a topic modeling algorithm, then click on a topic about film and television,  choose article about film director Stanley Kubrick which is associated with this topic, explore a
related topic about philosophy and psychology, and finally view a related article about Existentialism.
There have been numerous extensions for topic models framework. One specific example is the dynamic topic model \cite{blei2006dynamic} which allows the topic distribution to drift across time. 
Fig. \ref{fig:dynamicTM} shows two topics from a dynamic topic model fit to Science from 1880 to 2002. The topics at each decade are illustrated with the top words. Rather than a static summerization, we can use the dynamic topic model to track the topic's changes over time.

\begin{figure}[t!h!]
  \centering
    \includegraphics[width=.4\textwidth]{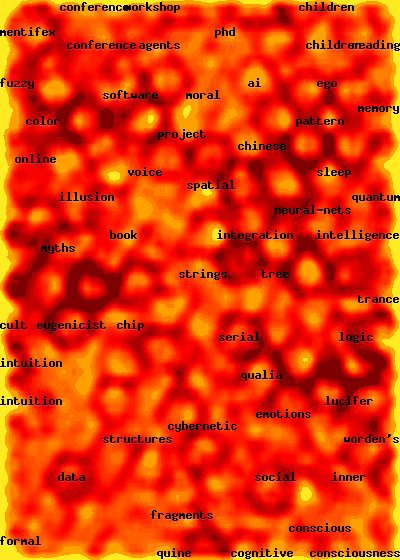} 
	\caption{WEBSOM text visualization system with SOM dimension reduction}
\label{fig:localrank}
\end{figure}

\begin{figure}[t!h!]
  \centering
    \includegraphics[width=.4\textwidth]{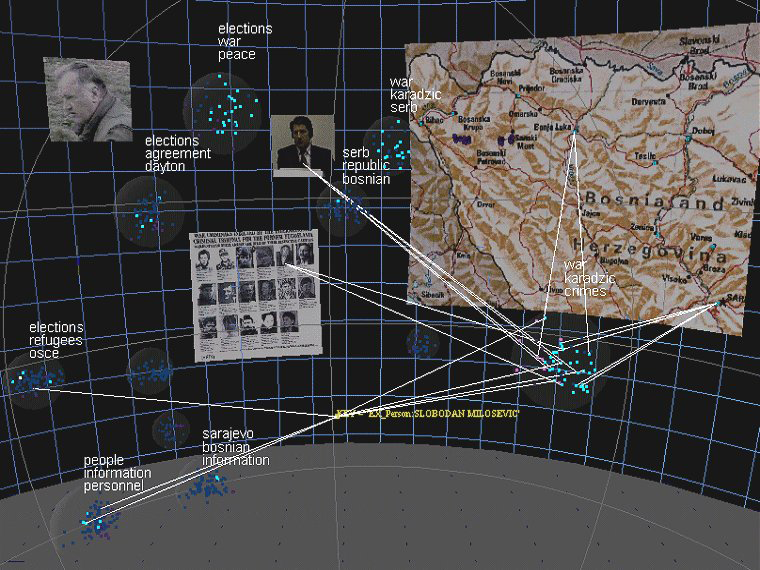} 
	\caption{Starlight's text visualization system using TRUST for dimension reduction}
\label{fig:starlight}
\end{figure}

For text analysis system built primarily for visualization, techniques that project the document from high feature space into 2 dimensional space is heavily used such as self-organizing map (SOM), Multidimensional scaling (MDS) as well as other specifically created projection algorithm
Lin's system \cite{lin1991self} in early 1990s demonstrate the use of self-organizing maps
for organizing text documents, where they used key index terms to extract vector space model for the documents, and train a SOM feature map.
WEBSOM \cite{honkela1997websom} was a later example for the SOM visualization. They
used a two-stage process that involved using an initial SOM to generate reduced
dimensionality text vectors that were then mapped with a second SOM for visualization
purposes.
. In a later work \cite{kohonen2000self} they extend teh system with the aim of scaling up the SOM algorithm to be able to deal
with large amounts of high-dimensional data.
 In a practical experiment
we mapped 6 840 568 patent abstracts onto a 1 002 240-node
SOM. As the feature vectors we used 500-dimensional vectors of
stochastic figures obtained as random projections of weighted
word histograms.
For MDS dimensional reducetion, an early example is the Bead system \cite{chalmers1992bead} developed during the early 1990's, which used document keywords and a hybrid MDS
algorithm based on an optimized form of simulated annealing to construct a vector space model.
IN-SPIRETM  \cite{wise1995visualizing} used multidimensional scaling, then anchored least stress, and
finally a hybrid clustering/principal components analysis projection scheme to map documents to a 2D space.
The text visualization system of Starlight's  \cite{risch1997starlight} uses the
Boeing Text Representation Using Subspace Transformation (TRUST) algorithm for dimension reduction and semantic labeling. Text vectors generated by TRUST are
clustered and the cluster centroids are down-projected to 2D and 3D using
a nonlinear manifold learning algorithm, as shown in Fig. \ref{fig:starlight}.


\section{Visualization for Model Construction}
Data mining is usually known as the highly automated model where the algorithm are designed with algorithmic or statistics techniques to do off the scene job. 
In recent years, however, marrying the classical techniques with the idea of info vis has stimulated the research on \textbf{Interactive model construction}, where visualization is provided 
for partial results of the current construction of model while a user can provide feedback to change the behavior of the process. 
The resulted approach will incorporate the computation power of the automatic analysis with the human wisdom from the user input, while at the same time greatly increase interpretability of the model.

\subsection{Progressive construction}
There is a large family of machine learning models that are built progressively. At each step, a "building block" is greedy selected to be a permanent part of the model. There are visualization techniques for these type of models to allow user to interfere with each step of the progressive construction.

A standard technique from machine learning community is the Visual decision tree construction.
 
from the machine learning platform of Weka \cite{ware2001interactive}.
The system enables the user to construct a decision tree graphically using the split method of bivariate splits. As shown in \cite{lubinsky1994classification}, this split method can generally achieve better performance and moreover it allow data to be interpreted visually in plots and tables.
Each bivariate split is represented as set of 2D polygons. Polygons are easy
to draw and can approximate arbitrarily complex two-dimensional shapes. 
In conjunction
with the standard recursive divide and conquer decision tree procedure, they
enable users to approximate the target concept to any degree of accuracy while minimizing
the number of splits that must be generated to identify pure regions of the instance
space.
They use two main module for the visualization,  tree visualizers and data detail visualizers.
At any stage the data at any given node in the tree 
can be visualized by left-clicking on that node. The process of interactively drawing polygons in instance space, defining splits and appending them to the tree continues until the user is
satisfied with the resulting classifier.

Another contribution of \cite{ware2001interactive} is the experiment they conduct comparing the automatic constructed model and user constructed model.They show that appropriate techniques can empower users to create models that compete with classifiers built by state-of-the-art learning
algorithms. They also found that success hinges on the domain: if a few attributes can support good
predictions, users generate accurate classifiers, whereas domains with many high-order
attribute interactions favour standard machine learning techniques.

\begin{figure}[t!h!]
  \centering
    \includegraphics[width=.4\textwidth]{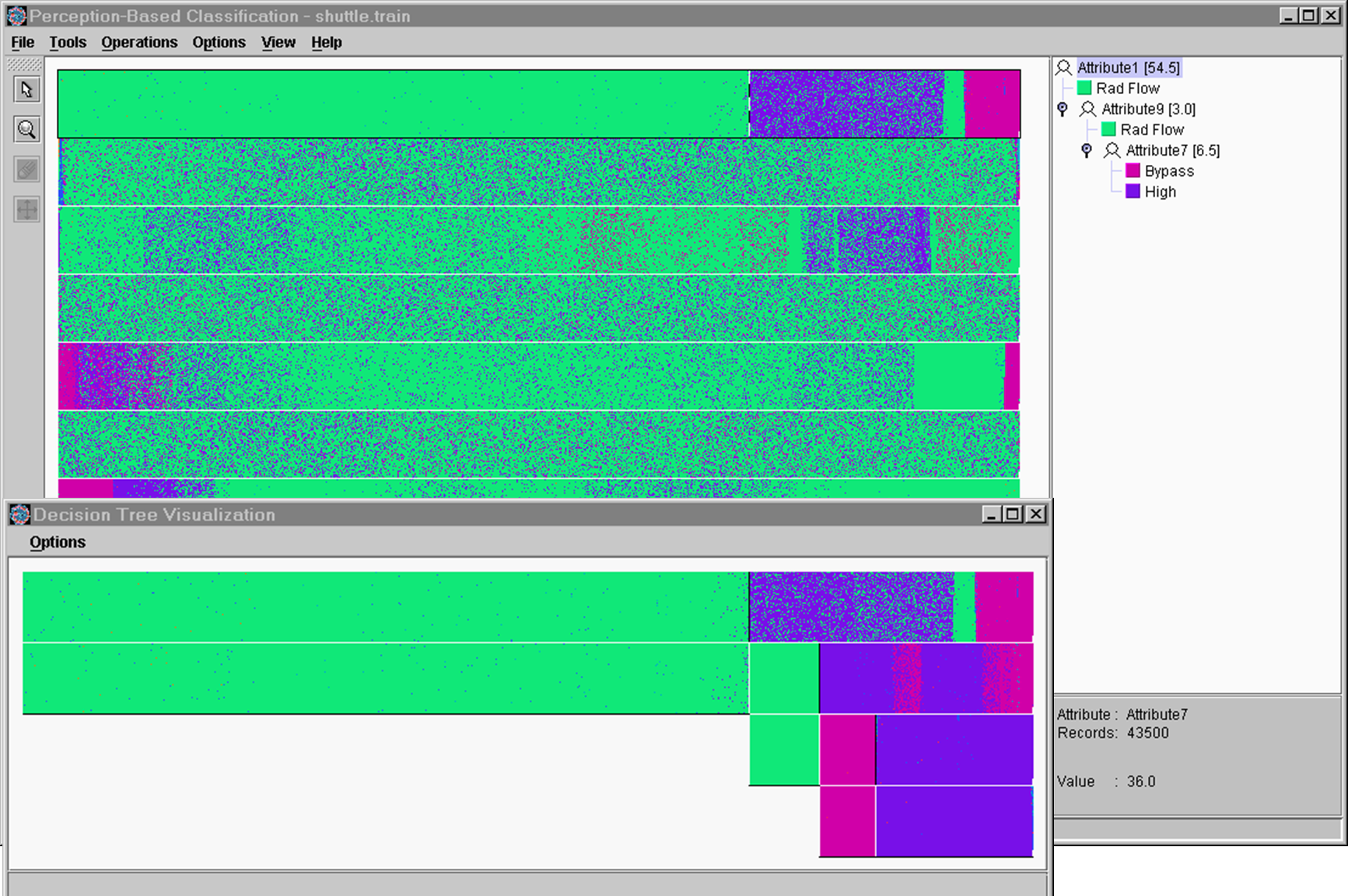} 
	\caption{the PBC system for interactive decision tree constructing.}
\label{fig:pbc}
\end{figure}

In \cite{ankerst1999visual} a more heavy weight interactive decision tree mechanism is introduced.
They proposed PBC -- Perception Based Classification, where different styles of cooperation
- ranging from completely manual over combined to completely
automatic classification are supported.
They propose two kind of encoding technique for the decision tree data where each attribute of the training data is visualized in a separate area of the plot, and different class labels of the training objects are represented by different colors.
The first is bar visualization. Within a bar, the
sorted attribute values are mapped to points line-by-line. Each attribute is visualized
independently from the other attributes in a separate bar, facilitatling the visualization of the hierarchical structure of decision tree nodes. 
The second is the tree visualization technique, each node is represented by the
data visualization of the chosen splitting attribute of that node. For
each level of the tree, a bar is drawn representing all nodes of this
level, which is then stacked together according to the hierarchical structure of the tree.
Fig. \ref{fig:pbc} shows the main interface of the model.
main window visualizes the bar visualization for data of the active node and depicts the
whole decision tree in standard representation. The additional window in the
foreground depicts the same decision tree using the new technique
of visualization.

\begin{figure}[t!h!]
  \centering
    \includegraphics[width=.4\textwidth]{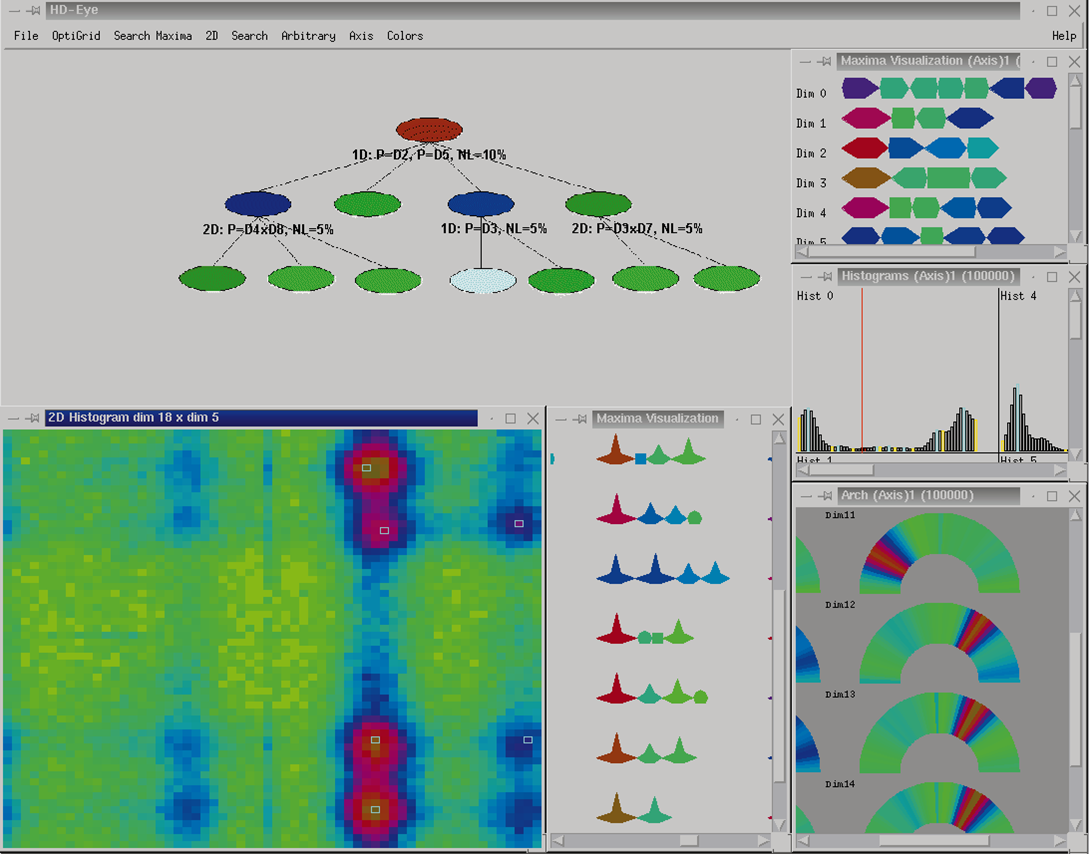} 
	\caption{the HDEye system for visual clustering.}
\label{fig:localrank}
\end{figure}

Such step-by-step construction paradigm is suitable for many kinds of progressive machine learning algorithms. Another type work that illustrates this tight coupling of visualization
resources into a mining technique is by Hinneburg et al.
\cite{hinneburg1999hd}. They describe an effective approach for clustering
high-dimensional data combining an advanced clustering
algorithm, called OptiGrid, with visualization methods that
support the interactive clustering process. The approach is a
recursive one: In each step, the actual data set is partitioned
into a number of subsets, if possible, and then the subsets
containing at least one cluster are dealt with recursively.
The partitioning in the framework of OptiGrid uses a generalized multidimensional grid defined by a
number of separators chosen in regions with minimal point
density. The recursion stops for a subset when no good
separators can be found. 
Choosing the contracting projections
and specifying the separators for building the multidimensional
grid, however, are two difficult problems that
cannot be done fully automatically because of the diverse
cluster characteristics of different data sets. 
Visualization technology can help in performing these
tasks.
Therefore, they developed a number
of new visualization techniques that represent the
important features of a large number of projections.
Finally they integrated all visualization techniques by using a
tree-like visualization of the projection and separator
hierarchy.

\subsection{Iterative prototyping}
Apart from machine learning models with explicit progressive structure, many models come with a more wholistic construction. Correspondingly, an alternative visualization strategy will be allowing user to give feedback to a visual prototype of the trained model, with emphasis on the the performance measure of the prototype, and effectively incorporate such user feedback to retrain the model. To fully exploit the advantage of interactiveness, such visualization system usually supports a iterative process where user keep reconstructing the model until satisfiable.

\begin{figure}[t!h!]
  \centering
    \includegraphics[width=.4\textwidth]{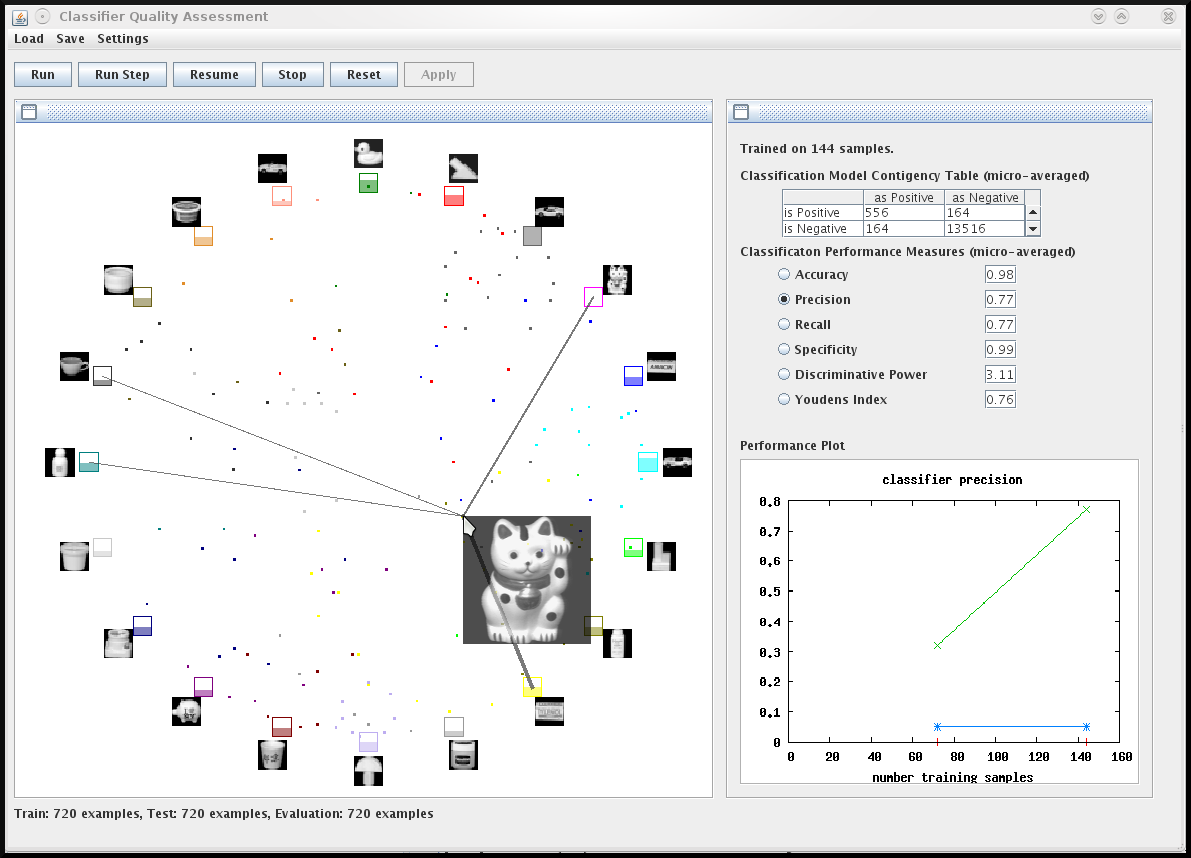} 
	\caption{The system by Seifert et.al. The classification probability view allows users to select a falsely  classified item to the right class and retrain the classifier.}
\label{fig:Seifert}
\end{figure}

The \cite{seifert2009visualization} present a visualization system to investigate and give feedback to all classifiers whose predictions can be interpreted as probability distribution over classes. Users can give feedback to classifiers directly in the visualization by drag and drop items to retrain the classifier with simple interaction techniques like drag and drop.
Figure \ref{fig:Seifert} shows the analysis scenario where the user is moving the item towards
the correct class “cat” . This item then serves as new training example for the particular class and the
classifier is re-trained and re-evaluated.

Such iterative refinement techinique can also be applied to regression models.
In \cite{lampeinteractive} they propose a new approach to rapidly draft, fit and quantify model prototypes in visualization space and demonstrate  that these models can provide important insights and accurate metrics about the original data. 
They conduct the visualization similarly to the statistical concept of de-trending. Data that behaves according to the model is de-emphasized, leaving outliers and potential model flaws which are critical for refining regression model. They also proposed a workflow for the refining process: visualize and observe,
sketch and fit, externalize and subtract, then iterate. A key step is the externalization, that transfers quantitative information back from
visualization space into model space.
The system is evaluated on streaming process data from the Norwegian oil and gas industry,
and on weather data, investigating the distribution of temperatures over the course of a year.

\begin{figure}[t!h!]
  \centering
    \includegraphics[width=.4\textwidth]{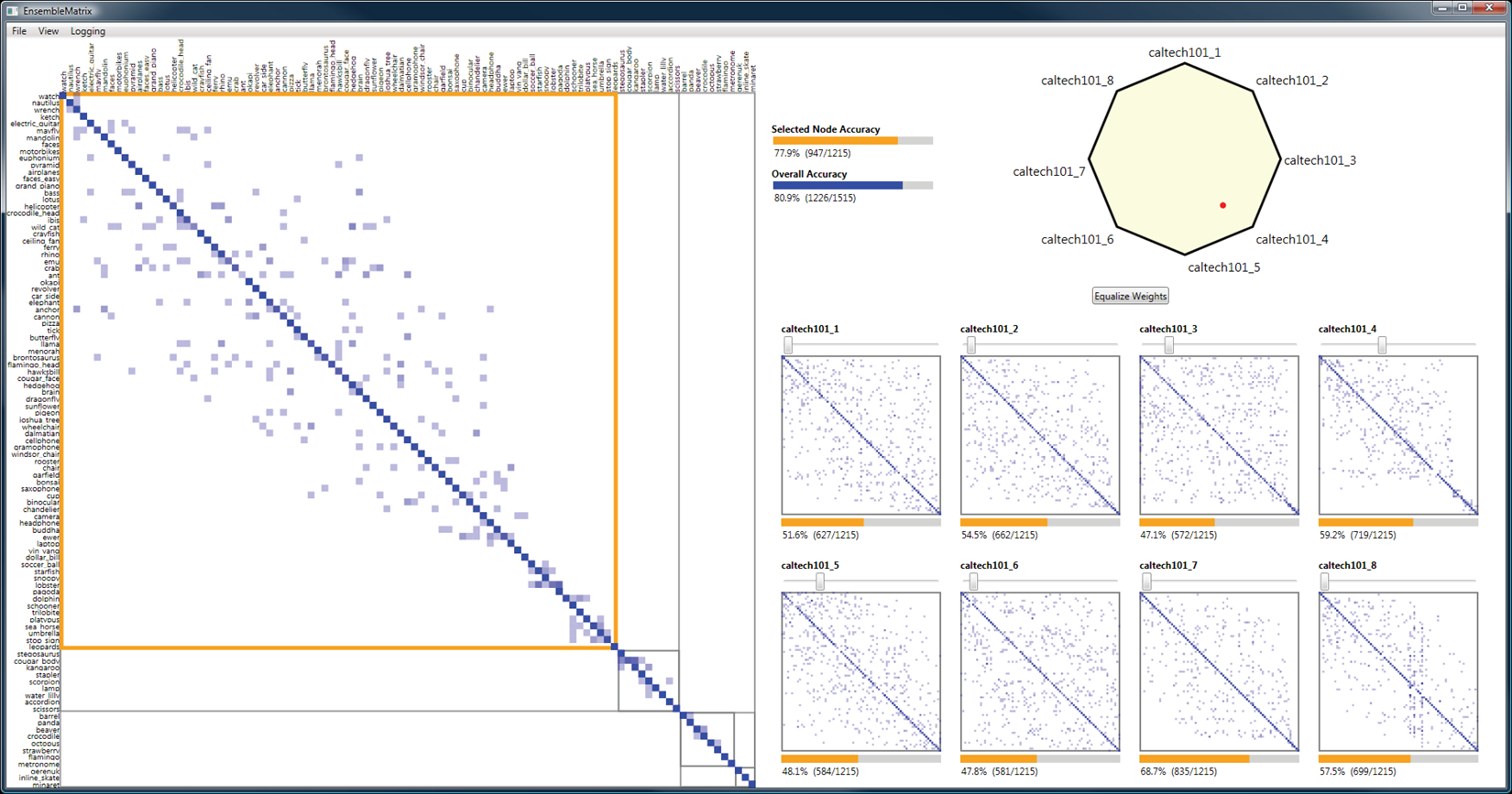} 
	\caption{The EnsembleMatrix system}
\label{fig:EnsembleMatrix}
\end{figure}
Another interesting aspect is in ensemble model construction. Ensemble is a common strategy that combine several different sub-model to reduce bias and improve precision \cite{dietterich2000ensemble}. Such kind of meta strategy provide the oppurtunity for interaction technique to rearrange the combination of sub components. 
The goal here is to select the best subsets of sub components that has good invidual performance while at the same time boost each other when performed jointly.
In \cite{talbot2009ensemblematrix}
they propose EnsembleMatrix,
an interactive visualization system that presents
a graphical view of confusion matrices to help users understand
relative merits of various classifiers by supplying a visual sum-
mary that spans multiple classifiers.

As shown in Fig. \ref{fig:EnsembleMatrix}, the EnsembleMatrix interface consists of three basic sections:
the Component Classifier view on the lower right,
which contains an entry for each classifier that the user has
imported to explore, the Linear Combination widget on the
upper right, and the main Ensemble Classifier view on the
left.
Confusion matrices of component classifiers are shown in thumbnails on the right. The
matrix on the left shows the confusion matrix of the current ensemble classifier built by the user.
EnsembleMatrix provides two basic mechanisms for users
to explore combinations of the classifiers. The first is a partitioning
operation, which divides the class space into multiple
partitions. The second is arbitrary linear combinations
of the Component Classifiers for each of these partitions
Experiment results show that users are able to quickly combine
multiple classifiers operating on multiple feature sets to
produce an ensemble classifier with accuracy that approaches
best-reported performance classifying images in
the CalTech-101 dataset.

\subsection{Interactive pipeline construction}
A even more algorithm agnostic approach for such interaction is treat each submodule of individual algorithm as black box with certain type of input and output and provide visualization system for users to assemble these submodules into a whole data mining pipeline.

\begin{figure}[t!h!]
  \centering
    \includegraphics[width=.4\textwidth]{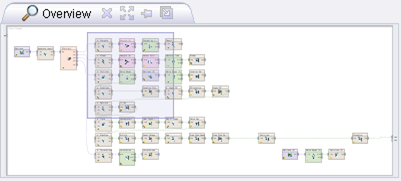} 
	\caption{The Overview Panel in Rapidminer system shows the entire data mining pipeline.}
\label{fig:localrank}
\end{figure}

Weka \cite{holmes1994weka} is a open source software for data mining purposes. The KnowledgeFlow panel presents a data-flow inspired interface allowing the user to create pipelines in order to perform data pre-processing,classification, regression, clustering, association rules, and visualisation. 
The interactive pipeline construction system provide users a intuitive data flow style layout that can 
process data in batches or incrementally, 
 process multiple batches or streams in parallel where each separate flow executes
in its own thread,
 chain filters together,
 view models produced by classifiers for each fold in a cross validation,
 visualize performance of incremental classifiers during processing by scrolling
plots of classification accuracy, RMS error, predictions etc.,
and plugin facility to allow easy addition of new components to the Knowl-
edgeFlow.
RapidMiner \cite{rapidminer2008rapidminer} is an enterprise software providing solutions for a wide range of analytic tasks including banking, insurance, retail, manufacturing, oil and scientific research.
allows the user to create data flows, including data import, pre-processing, execution process and visualisation. 
Similar approach is also adopted by KNIME \cite{berthold2008knime}, STATISTICA \cite{statistica2013statistica},  Orange \cite{demvsar2004orange},  
as well as many other practical data mining systems\cite{kagdi2007survey, goebel1999survey}.


\section{Visualization for Model Evaluation}
The end product of data mining usually involves a very large set of complex decision rules, together with sophisticated performance measure on the benchmark test data. Different families of techniques are widely applied for \textbf{visualization of model evaluation}. 
By visually conveying the results of a mining task, such as classification, clustering or other form of data mining, these visualization approaches will better answer the question of what the conclusions are, why a model makes such conclusions, and how good they perform, which is very critical for human decision makers to analyze the behavior of model, apply the model output, and reflect on the results.

\subsection{Classification}
The task of classification refers to the problem of identifying category membership for new observation , on the basis of a training set of data containing observations whose category membership is known.
Visualization for such form of model typically takes the input of data with various features, the predicted classes of each item as well as the ground truth. The visualization may also include model specific structures, such as the decision boundary for classifiers.

\subsubsection{Test oriented visualization}
A number of measurements 
have been developed to address the question of how well the model performs on the benchmark test data and compare the prediction of model with ground truth. 
Although simple statistics of such measurements like precision, recall, specificity, F-measure and confusion matrix\cite{powers2011evaluation} has been proposed, the "behind the scene" performance for different part of data as well as the confidence for each prediction are usually left unexamined by such measure. 
Visualization of such measurements aims to overcome these challenges and provide deeper understanding of model conclusions for both right and wrong predictions.

Common way of graphically tuning, assessing, and comparing the performance of classifiers within the community of classical machine learning and statistics include the family of receiver operating characteristic (ROC) curves \cite{fawcett2006introduction}. Instead of simply looking at the prediction label, these models will in addition inspect the confidence by plotting the true positive rate against the false positive rate at various probability threshold. The extensions including such as cost curve that incorporates misclassification costs \cite{drummond2006cost}, three-way ROC and Multi-class ROC for classification with multiple classes \cite{hassan2010novel}, and regression error characteristic (REC) Curves and the Regression ROC (RROC) curves for regression problems \cite{HernandezOrallo20133395}.

\begin{figure*}[t!h!]
  \centering
    \includegraphics[width=.9\textwidth]{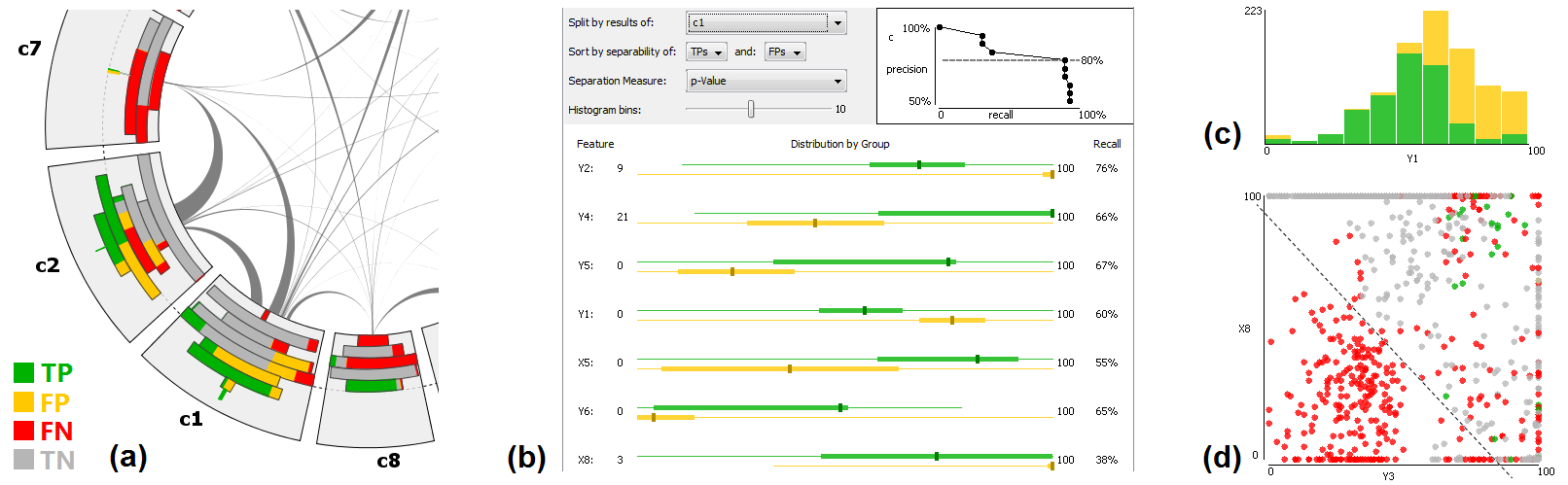} 
	\caption{Alsallakh et.al's tool, showing (a) the confusion wheel (b) the feature analysis view  (c, d) histograms and scatterplots for the separability of the selected features}
\label{fig:Alsallakh}
\end{figure*}

As the field field of machine learning and data mining become eminent, more heavy weighted visualization system for such purpose has been introduced.
\cite{alsallakh2014visual} propose a comprehensive visualization system for posterori classification results analysis to provide insight into different aspects of the classification results for a large number of samples.
One visualization emphasizes at which probabilities these
samples were classified and how these probabilities correlate with classification error in terms of false positives and false negatives.
As shown in Fig. \ref{fig:Alsallakh}, they employ the visual layout of Contingency Wheel++ \cite{alsallakh2012reinventing} which places these histograms in a ring chart whose sectors represent different classes.
In contrast to the matrix, this layout emphasizes
the classes as primary visual objects with all information related
to a class grouped in one place.
For each class, the samples S are divided into four sets according
to their classification results:TP, FP, TN, and FN, further divided by confidence.
The chords between the sectors encode the confusion wheel depicts class confusions
. A chord between two classes is depicted with a
varying thickness denoting the confusion rate for each corresponding pair.
The feature view on the right emphasizes the features of these samples and ranks them by their separation power between selected true and false classifications where they use techniques such as
Boxplots, stacked histogram, separation measures, recall-precision graph for 
analyzing how the data features are
distributed among the affected samples.
The system also allow interactively defining and evaluating post-classification rules.

Ensembles for different classifiers are a common way to boost the performance for weaker classifiers.
It is widely adopted because its superior performance in
comparison to single classifiers methods both in theory and practise. Visualizing the performance of different sub classifiers as well as the ensemble classifier is a key component for inteprating the structure of ensemble methods.
\cite{stiglic2006using} provide a simple tool to display different sub decision trees as well as the ensemble one with graphical intepretation and concrete decision rules, basic zooming and navigation is provided.
In \cite{urbanek2002exploring}
they describes different
views and methods to analyze decision forests  
including fluctuation diagram, parallel coordinates
plots where the deviance gain for specific variable in different sub-tree is emphasized.

\subsubsection{Instance space based visualization}
 
A common approach to visually characterize a classifer is to focus on input data instance and how the model behave on different region of the instance space. 
Since the dimensionality of instance of data are usually high, visualization of such data usually take different dimensionality reduction approach to navigate the viewer through the high-dimensional spaces
and characterize the data instance and classification results.
In addition, for models that based its conclusion from clues in the instance space, such as maximal margin separating hyperplane for SVM, we can also further incorporate these specific structure into the instance space visualization.

\begin{figure}[t!h!]
  \centering
    \includegraphics[width=.4\textwidth]{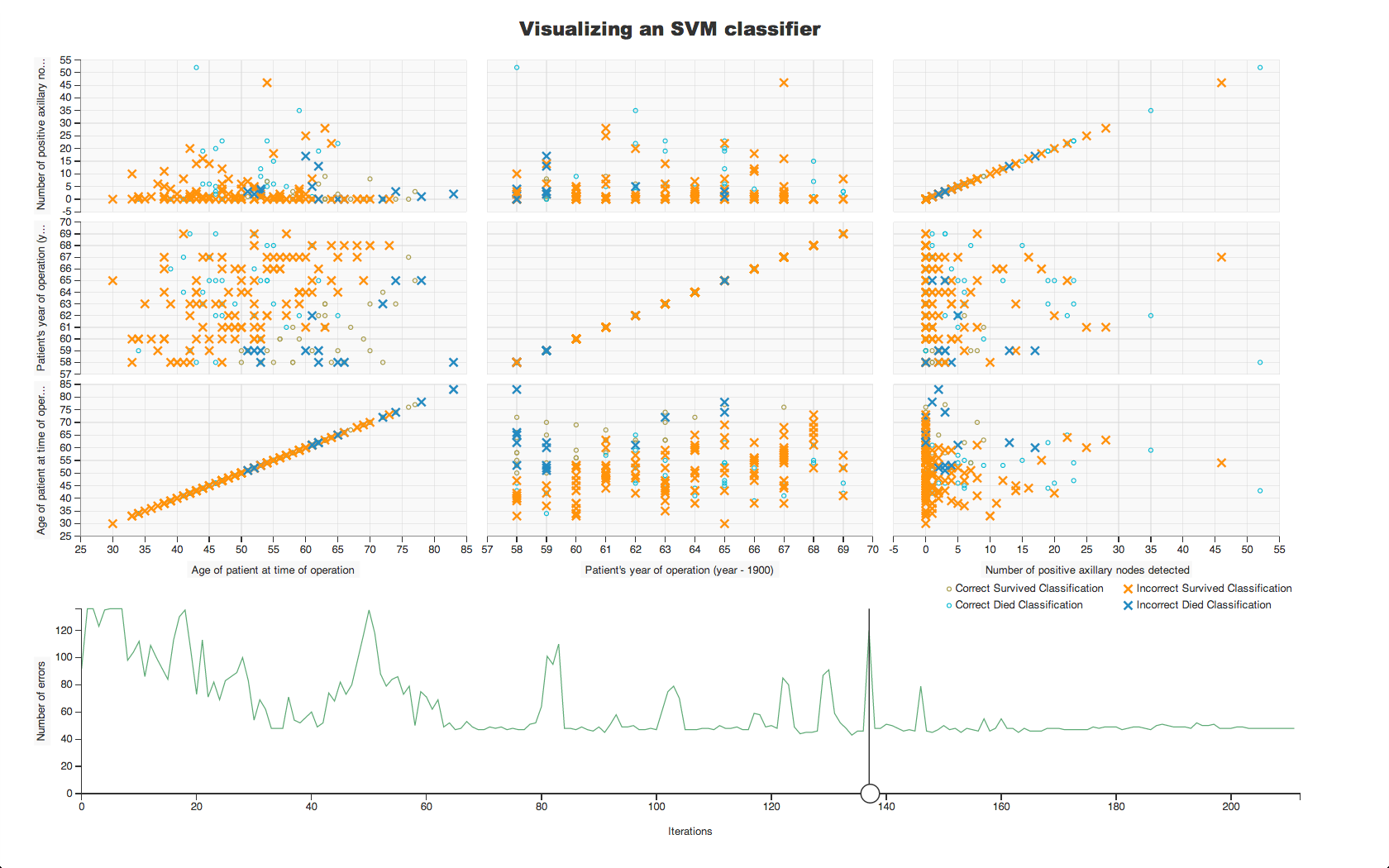}
	\caption{Byklov et. al's tool showing showing the scatterplot matrix for an iteration}
\label{fig:Byklov}
\end{figure}

One way to deal with high dimensional data instance is to comprehensively show the pairwise feature projection for the data using techniques such as SPLOM.
Bykov et.al \cite{bykov2009interaction} 
 create a SPLOM based visualization tool for analyzing
machine learning classification algorithms
by providing the user with per-iteration
performance information for the algorithm.
This is done
through two main views. The first view contains a
scatterplot matrix of the data projected into multiple
dimension pairs. Each point is labeled with its actual and
predicted labels to highlight where in the dataset errors
occur at each dimension. The second view provides
summary statistics (classification accuracy, number of
errors, etc.) at each iteration and an interface to scroll
through every iteration of the algorithm.
Fig. \ref{fig:Byklov} shows the main screen of the visualization. 
where user move the slider on the x-axis of the bottom
chart to switch the scatterplot matrix to a specific different iteration  in which the number of errors sharply increased. 
.
The scatterplot on the top shows different features plotted against each other, with each kind classification outcome encoded with different point glyph.
 The sudden increase in errors can
easily be seen by the sudden increase in orange and blue marks representing different kind of classification errors.

\begin{figure}[t!h!]
  \centering
    \includegraphics[width=.4\textwidth]{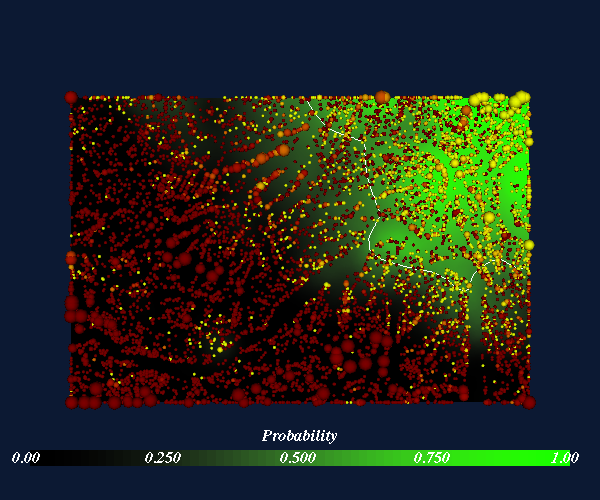} 
	\caption{Rheingans et.al's tool, showing SOM probability map with test instances}
\label{fig:Rheingans}
\end{figure}

A very common way to deal with high dimensionity is non linear dimension reduction.
Such dimension reduction techniques capture the feature of data item using a specific location, 
revealing the spatial relation of both positive and negative prediction. More importantly, it facilitates the visualization of classifier stucture such as decision boundary. 
Rheingans et.al introduce the \cite{rheingans2000visualizing} techniques for high-dimensional data
space projection, display of probabilistic predictions, variable/class
correlation, and instance mapping.
They used a set of projection techniques based on self-organizing maps (SOM)
. Figure \ref{fig:Rheingans} shows a representation of the model
where the data space has been projected to two dimensions using
a SOM. 
The decision boundary, i.e. the boundary between positive
and negative predictions projected on the 2D plane is shown by a white line. 
Glyph size indicates the number of instances at a given
point.  A continuous color
map is used to show the proportion of class labels in the set of collocated
instances. Yellow shows predominantly positive instances,
red shows predominantly negative instances.
Test instances is encoded as a small sphere-shaped glyph, with size indicating the density of instances 
and a continuous color indicating the proportion of class labels in the set of collocated, yellow positive for and red for negative. Notice that in the plot many of the misclassified instances are located in the vicinity of decision boundary, showing they are not too far from "right".
Frank et.al\cite{frank2003visualizing} provid similar technique for instance space projection 
to visualize class probability estimates for different regions of the instance space. They discuss different choices for kernel density estimator and sampling strategy for visualizing the expected class probabilities in the projected space and render the class probabilities using RGB color encoding.
In \cite{caragea2001gaining} they explore the use of
the projection methods of tours \cite{elmqvist2008rolling,dianne1997m} to exam-
ine results from SVM, 
where they use the grand  tour for generally exploring sup-
port vectors and classification boundaries,
manually controlled tours for studying variable importance, and correlation tour for examining predicted values in relation to explanatory variables.

\subsubsection{Factor analysis}
The process of generating conclusions for classifer usually involve different factors from the feature of data instance.  Visualizing the relationship between the different factors and the final conclusions for different data instance will shed light on the question of why a model makes such conclusions.

\begin{figure}[t!h!]
  \centering
    \includegraphics[width=.4\textwidth]{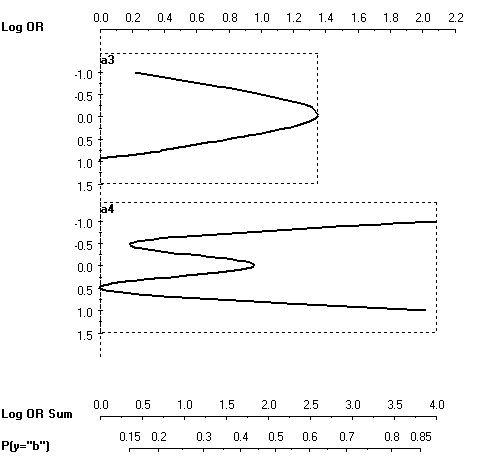} 
	\caption{Nomograph showing the nonlinearities in the ionosphere data set for SVM with RBF basis}
\end{figure}

Nomograms
are an established model visualization technique that can
graphically encode the complete model on a single page.
The dimensionality of the visualization does not depend on
the number of attributes, but merely on the properties of
the kernel. By providing an easy-to-interpret visualization
the analysts can gain insight and study the effects of predictive
factors.
has also been applied to various models including logistic regression \cite{lubsen1978practical}, Nayes Bayesian Classifier \cite{movzina2004nomograms} and support vector machines \cite{jakulin2005nomograms} to reveal 
the structure of the model and the relative influences of the attribute values
to the class probability.
As an example, \cite{jakulin2005nomograms} propose a simple yet potentially very effective way of
visualizing trained support vector machines. 
 To represent the effect of each predictive feature
on the log odds ratio scale as required for the nomograms,
we employ logistic regression to convert the distance from
the separating hyperplane into a probability. 
similar technique 

The CViz\cite{han2000ruleviz} system provide an alternative approach of extending the static parrellel coordinate view for multivariate data visualization.
They describe 
an interactive system for visualizing the process of
classification rule induction,  
encoding the data preprocessing and classification rules upon the parrellel coordinate platform.
The original data is visualized
using parallel coordinates and the user can see the results of
data reduction and attribute discretization on the parallel
coordinate representation. The discovered classification
rules are also displayed on the parallel coordinates plots
as rule polygons, colored strips as depicted in Fig. 11, where
a polygon covers the area that connects the (discretized)
attribute values that define particular rules. Rule accuracy
and quality may be coded by coloring the rule polygon and
user interaction is supported to allow focusing on subsets of
interesting rules.

\subsection{Clustering}
Other than classification, there are a number of unsupervised learning method widely used to discover hidden structure of the data to gain insight and facilitate decision making. 
Clustering is a specifically useful technique of rearranging the dataset and grouping similar items together, resulting in different clusters which are easier to analyze.
Visualization technique for clustering usually directly take the instance based approach, explicitly rendering invidual instance or aggregate of some instance as well as the inclusion relation between cluster and instances.

\begin{figure}[t!h!]
  \centering
    \includegraphics[width=.4\textwidth]{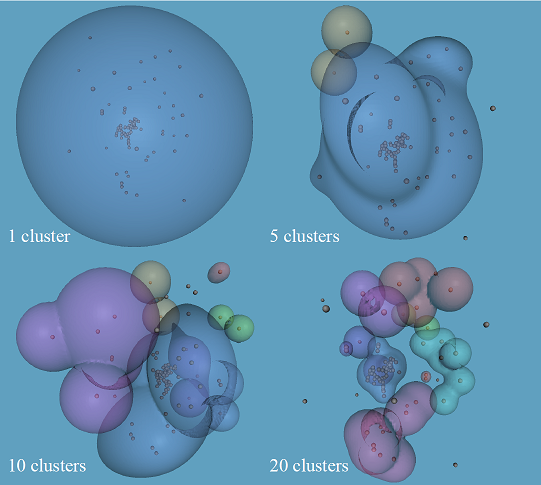} 
	\caption{The H-BLOB visualizes the clusters by computing a hierarchy of implicit surfaces}
\label{fig:localrank}
\end{figure}

One good example for the rendering of feature space, the instance and the inclusion relation as well as the internal hierarchical structure across the different clusters is 
is the BLOB and H-BLOB clustering visualization \cite{sprenger2000h}
, which use implicit surfaces
for visualizing data clusters.
BLOB\cite{gross1997visualizing} explicitly represent clusters by exhibiting
them in an enclosing surface
 that approximates the outline of their
included data objects as closely as possible. This is done by 
superimposing all field functions
in space and accordingly run a marching cube
algorithm \cite{98399} to extract the implicit surface at a given isovalue.
The later work H-BLOB further 
develop computation method for a hierarchy of implicit surfaces to visualize the hierarchical structure clustering of cluster tree, using the idea of computing higher level cluster based on child clusters in the cluster tree.

\begin{figure}[t!h!]
  \centering
    \includegraphics[width=.4\textwidth]{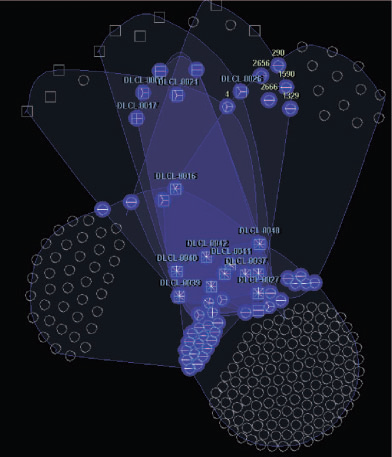} 
	\caption{Bicluster visualization, showing the overlapping, similarity and separation for clusters and entities.}
\label{fig:Bicluster}
\end{figure}
There are method tailored for different variation of clustering method.
BicOverlapper\cite{santamaria2008bicoverlapper} is a framework to visualize biclustering, which group the items under a certain subgroup of conditions and allow overlapping between clusters.
In order to improve the visualisation of biclusters, a
visualisation technique (Overlapper) is proposed to simultaneously represent
all biclusters from one or more biclustering algorithms by means of intersecting hulls, based on a force directed layout.
The use of glyphs on gene and conditions nodes
improves our understanding of instances of overlapping when
the representation becomes complex. 

Fig. \ref{fig:Bicluster} shows the bicluster analysis of a microarray data matrix
containing two types of Diffuse Large B-Cell Lymphomas \cite{1250903} achieve by the OPSM biclustering method.
Biclusters grouping mainly conditions or genes are easily identified, revealing asymmetry in OPSM method. The relaxed condition of order preservation searched by OPSM produces very large biclusters in some cases. Conditions grouped in all the
biclusters of OPSM have a strong influence in order preserving of gene expression levels. Most of them correspond to activated B-like lymphomas.

\begin{figure}[t!h!]
  \centering
    \includegraphics[width=.4\textwidth]{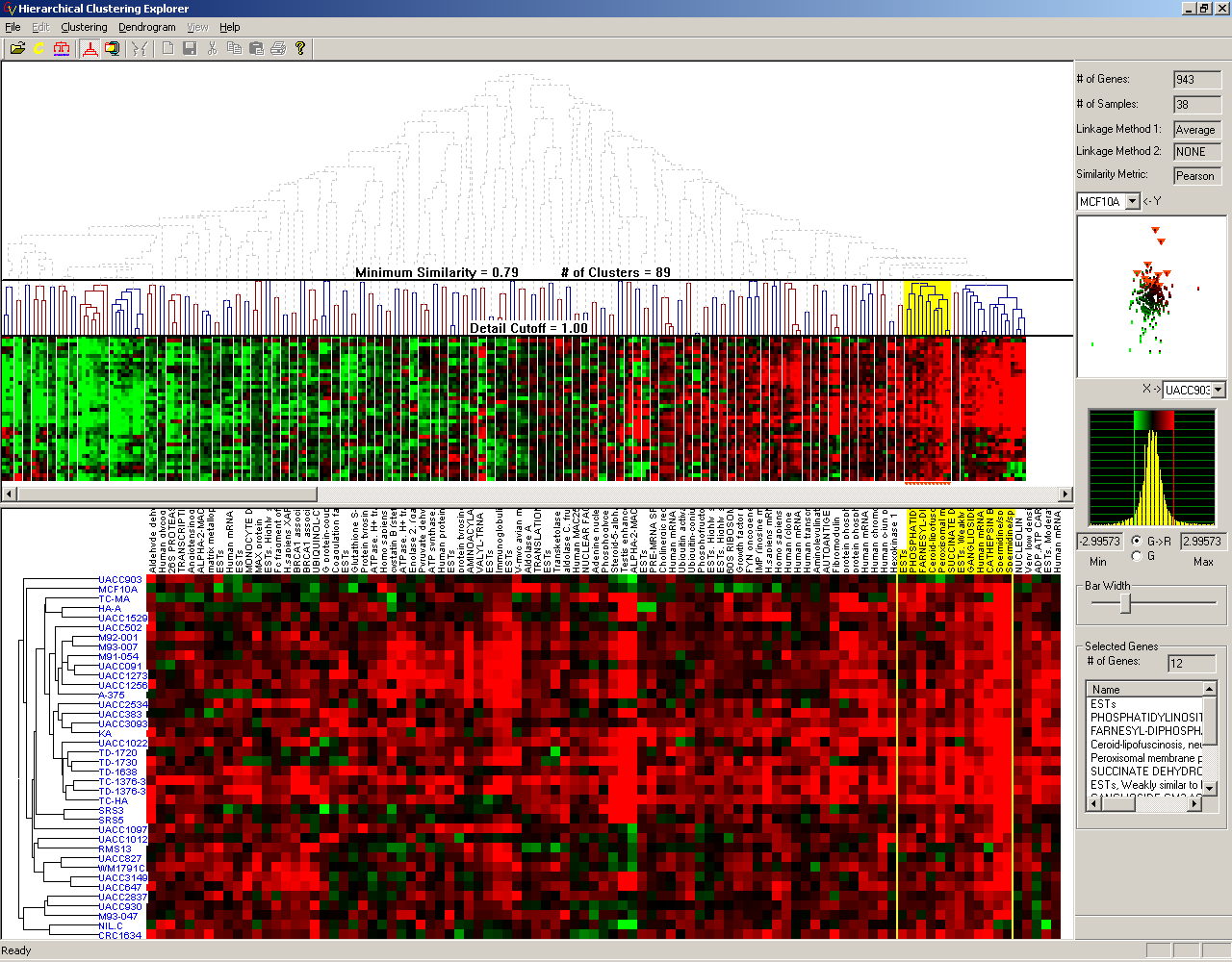} 
	\caption{HCE }
\label{fig:HCE}
\end{figure}

Instead of rendering the instance space, some visualization show the instance in aggreate abstract from.
The system of HCE(Hierarchical Clustering Explorer) \cite{seo2003understanding} focuses on hierarchy structure across different clusters and together with the interplay between other hierarchy structures.
Fig. \ref{fig:HCE} shows a screenshot of the interface. Each cluster is easily identified by the alternating
colored lines and the white gaps between clusters. The detail
information of a selected cluster yellow highlight in upper left is provided below the
overview together with the gene names and the other cluster dendrogram. 
It also allows interactive visual feedback (dendrogram and colour mosaic) and dynamic query
controls for users to refine the natural grouping.

Clustering can be further combined with domain knowledge for specific analytics application. As an example, 
Cadez et al.\cite{cadez2000visualization} describe an clustering based approach for  analysis and visualization of the dynamic behavior of visitors of a particular Web site. 
They focus on clustering users with similar behavior and then perform analysis on users behavior for each cluster.
Their visualization tool uses multiple windows to display user data regarding the multiple clusters. Sequences of rows within a window show the paths of single users through the site, each path being color coded by category reflect the different types of service provided by the site . summary information about clusters are also provided. The tool can help site administrators in identifying navigation patterns that may actually suggest actions to be taken to improve the site.

\subsection{Association Rules}

\begin{figure}[t!h!]
  \centering
\subfigure[Double-Decker plots]{
   \includegraphics[width=.2\textwidth] {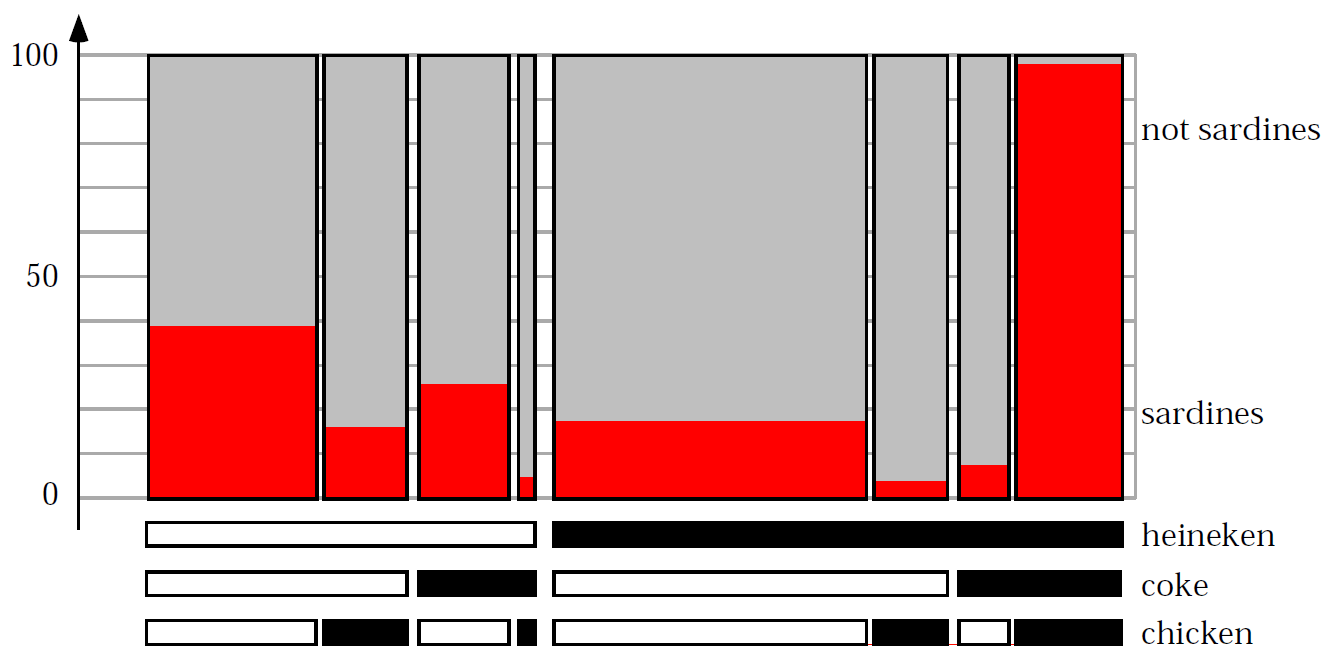}
   \label{fig:subfig1}
 }
  \subfigure[Parallel Coordinates]{
   \includegraphics[width=.2\textwidth] {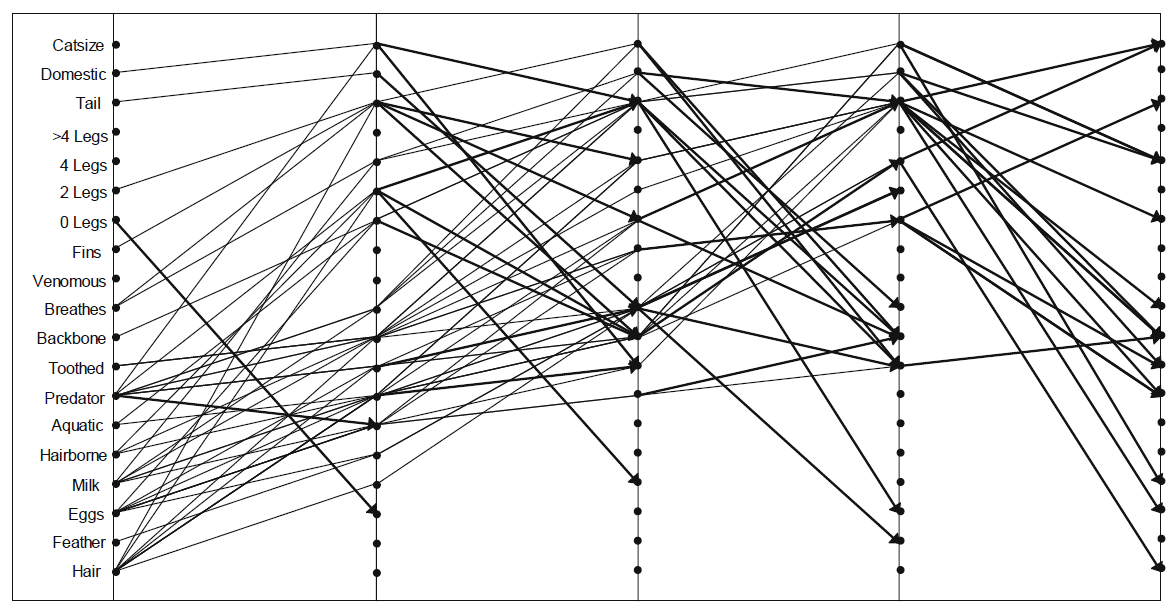}
   \label{fig:subfig2}
 }
\subfigure[Network]{
   \includegraphics[width=.2\textwidth] {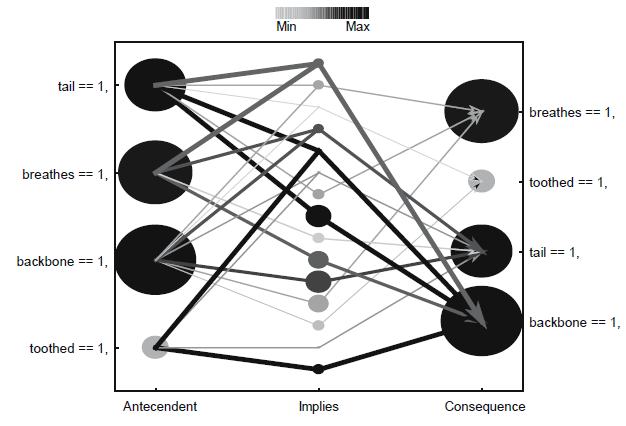}
   \label{fig:subfig2}
 }
 \subfigure[TwoKey plot]{
   \includegraphics[width=.2\textwidth] {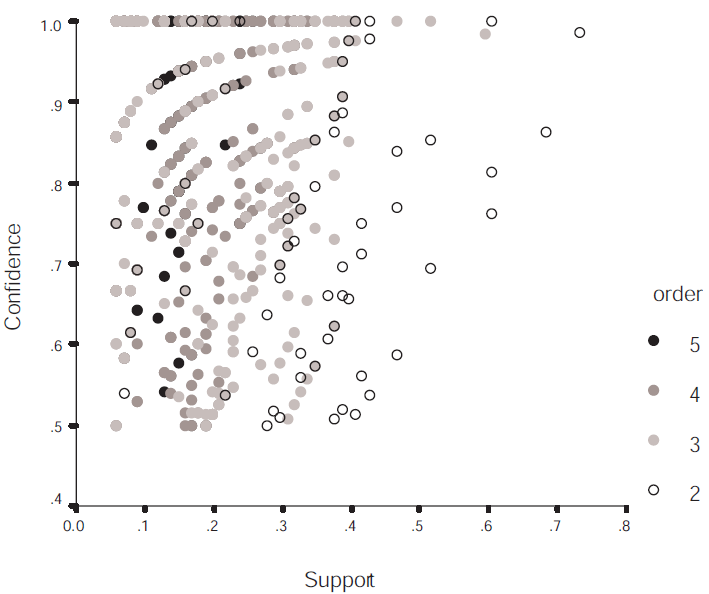}
   \label{fig:subfig2}
 }

	\caption{Different ways of  Association Rules  visualization}
\label{fig:localrank}
\end{figure}

The Association Rules (AR) is another fundamentally different data mining technique.
 It takes the input of a large database, and  provide rules describing 
interdependence between different items, each 
composed of the antecedent and the consequence with adequate amount of support and confidence. \cite{agrawal1993mining}
Association Rules are one of the most widespread data mining tools because they can be easily mined, even from very huge database. 
 
However, the results of mining are usually large quantities of rules, which 
require a massive effort in order to make actionable the retained knowledge. 
A main challenge for these visualization system is how to concisely encode each rule to accomodate hundreds of even more rules in one plot.

Double-Decker plots \cite{hofmann2000visualizing} 
 provide a visualization for single association rules but also for all its the
related rules. They were introduced to visualize each element of a multivariate
contingency table as a tile in the plot and they have been adapted to
visualize all the attributes involved in a rule by drawing a bar chart for the
consequence item and using linking highlighting for the antecedent items.

Different approaches has been developed for encoding association rules as a network \cite{cabena1999intelligent,statsoft1995statistica}.
In \cite{statsoft1995statistica}, a network representation of AR is provided where 
item and rules are encoded as nodes and each rule node is linked to corresponding items in antecedent and consequence by edges.
. The support
values for the antecedent and consequence of each association rule are indicated
by the sizes and colours of each circle. The thickness of each line indicates the
confidence value while the sizes and colours of the circles in the center, above the
Implies label, indicate the support of each rule.

The TwoKey plot \cite{unwin2001twokey} is a 2D plot approach for AR visualization that represents the rules according to their confidence and
support values. In such a plot, each rule is a point in a 2-D space where the xaxis
and the y-axis ranges respectively from the minimum to the maximum values
of the supports and of the confidences and different colors are used to highlight
the order of the rules.

Parallel coordinates have also been used for AR visualize to deal with the high dimensionality of the rules \cite{bruzzese2003parallel, kopanakis2001visual, yang2003visualizing}. 
For example, the approach proposed by \cite{yang2003visualizing} starts from arranging items by
groups on a number of parallel axes. A rule is represented as a polyline joining the items in the antecedent followed by an arrow connecting another polyline for the items in the consequence.

\subsection{Neuron Network}
Neuron Network are specific form of computing model inspired by biological neural networks that can be used for classification, regression or unsupervised learning. 
Being the state of the art approach on many pattern recognition or machine learning image classification problems, it has drawn specific attention from visualization community \cite{krizhevsky2012imagenet}.
The model of Neuron Network is composed of one or more layers of  interconnected "neurons" which can compute values from inputs.

\begin{figure}[t!h!]
  \centering
    \includegraphics[width=.4\textwidth]{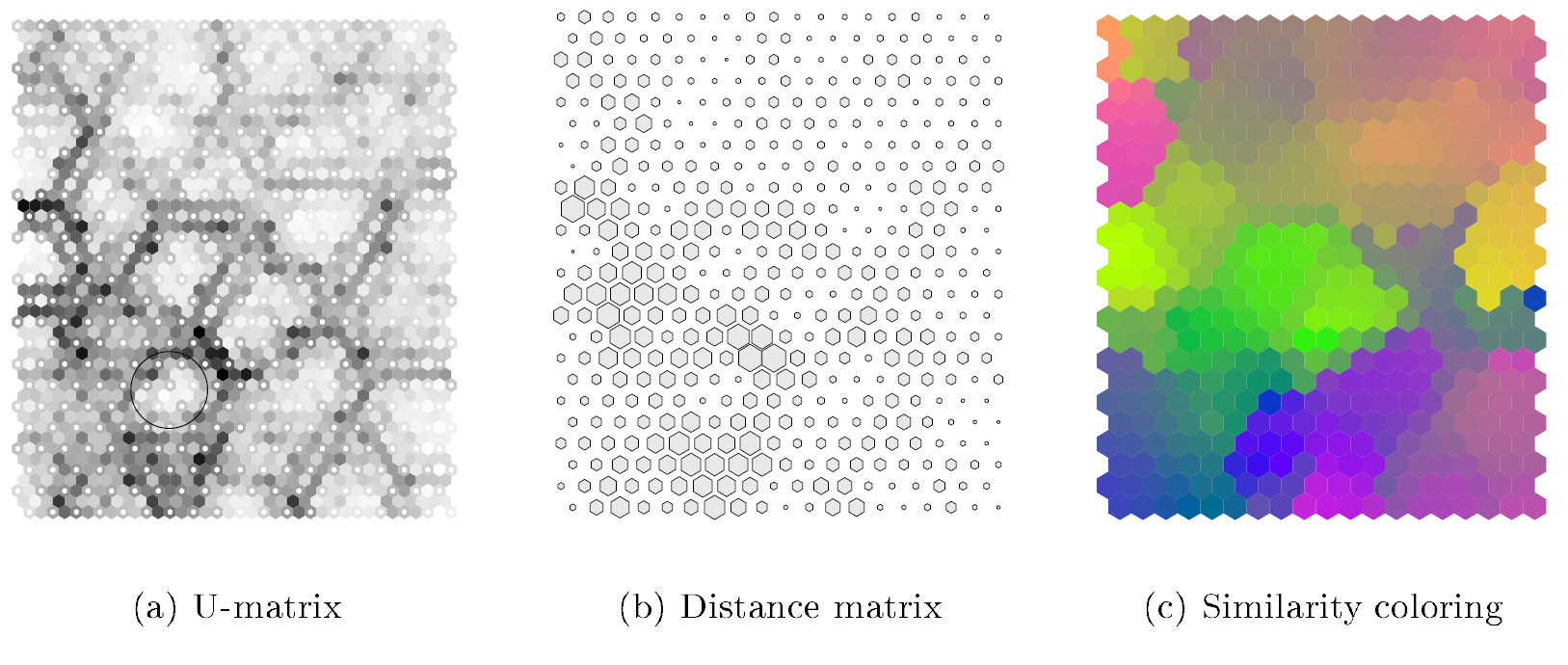} 
	\caption{Different techniques for showing cluster structure among SOM cells.}
\label{fig:som}
\end{figure}

The SOM is a neural network
algorithm for vector
quantization and projection that implements an
ordered dimensionality reducing mapping which follows
the probability density function of the training data. 
The SOM works by iteratively 
At each iteration in the training, the algorithm select a data sample, find the its best matching unit (BMU), and move the prototype vectors of the BMU as well as its neighbors in the grid towards the sample.
The resulted neurons grid can be easily arranged in 2D, with neighboring locations in the display space correspond
to neighboring locations in the data space.
\cite{vesanto1999som} provide a comprehensive survey of visualization techniques for SOM, where they identify three main categories, the methods for revealing the shape and structure, the methods for analyzing prototype vectors, and the methods examining new data samples.
Fig. \ref{fig:som} shows several common ways to visualize the cluster structure using distance matrix techniques. Figure (a) shows the U-matrix visualziation, where white dots represents the neurons and hexagon represents the values of U-matrix. The gray scale encodes the distance to the neighboring unit. Darker color means bigger distance. Clusters in the plot can be observed as light areas enclosed by darker borders. Figure (b) use size of map unit to encode the average distance to its neighbors. Figure (c) takes yet another approach by encoding the closeness in input space by similarity in the hue of colors for each hexagon area.

\begin{figure*}[t!h!]
  \centering
    \includegraphics[width=.9\textwidth]{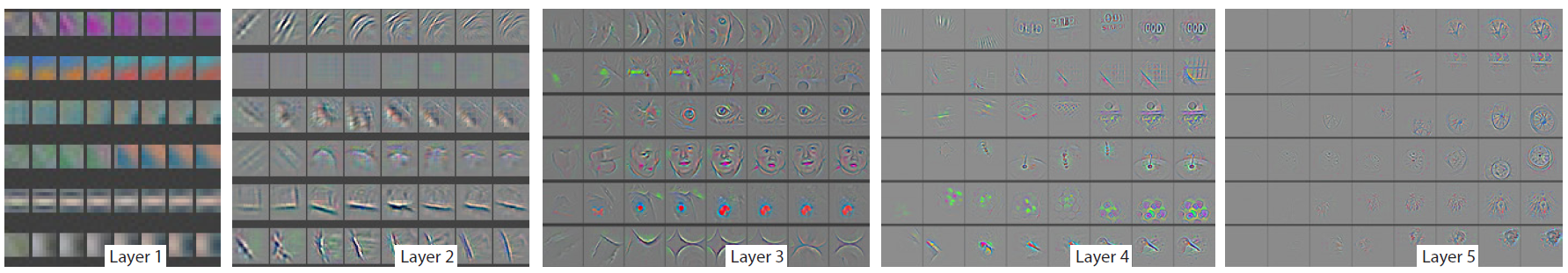} 
	\caption{Visualization of neurons shows evolution of a randomly chosen neurons in different layers through the training process. Each layer's features are displayed in a different horizontal block.}
\label{fig:convNN}
\end{figure*}

The latent semantic of neurons in multilayer network is usually captured by the input image with the strongest activation, either through numerical observation or selecting from training examples, which is the crucial component for visualizing multilayer complex neuron network.
\cite{zeiler2013visualizing,deepviz} is an example of using such neuron visualization technique 
to give insight into the function of intermediate
feature layers and the operation of
the classifier. 
Fig. \ref{fig:convNN} visualizes
the progression during training of the strongest
activation across all training examples within a given
feature map projected back to pixel space. Sudden
jumps in appearance result from a change in the image
from which the strongest activation originates. The
lower layers of the model can be seen to converge
within a few epochs. However, the upper layers only
develop develop after a considerable number of epochs
(40-50), demonstrating the need to let the models train
until fully converged.
Used in a diagnostic role, these
visualizations allow us to find model architectures
that outperform Krizhevsky et al. on
the ImageNet classification benchmark.

\begin{figure}[t!h!]
  \centering
    \includegraphics[width=.4\textwidth]{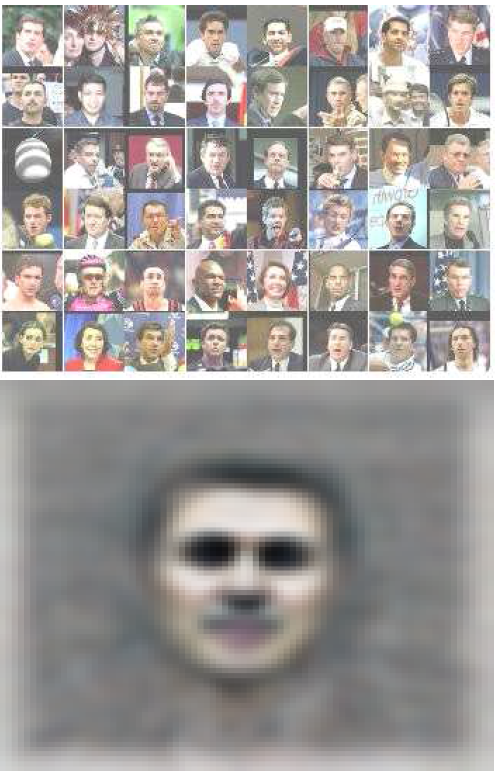} 
	\caption{Deep learning for Building High-level Features}
\label{fig:deepl}
\end{figure}

\cite{le2013building} extend this approach to sematic conceptual features.
uses Deep learning 
to address the problem of building highlevel,
class-specific feature detectors from
only unlabeled data. 
They present two visual techniques to verify if the optimal stimulus of the neuron is
indeed a face. The first method is visualizing the most
responsive stimuli in the test set. Since the test set
is large, this method can reliably detect near optimal
stimuli of the tested neuron. The second approach
is to perform numerical optimization to find the optimal
stimulus
They train a 9-
layered locally connected sparse autoencoder
with pooling and local contrast normalization
on a large dataset of images
for Building High-level Features. Fig. \ref{fig:deepl} show
top 48 stimuli of the best neuron from the test set as well as The optimal stimulus according to numerical
constraint optimization

\subsection{Graphical Probabilistic Models}

\begin{figure*}[t!h!]
  \centering
    \includegraphics[width=.9\textwidth]{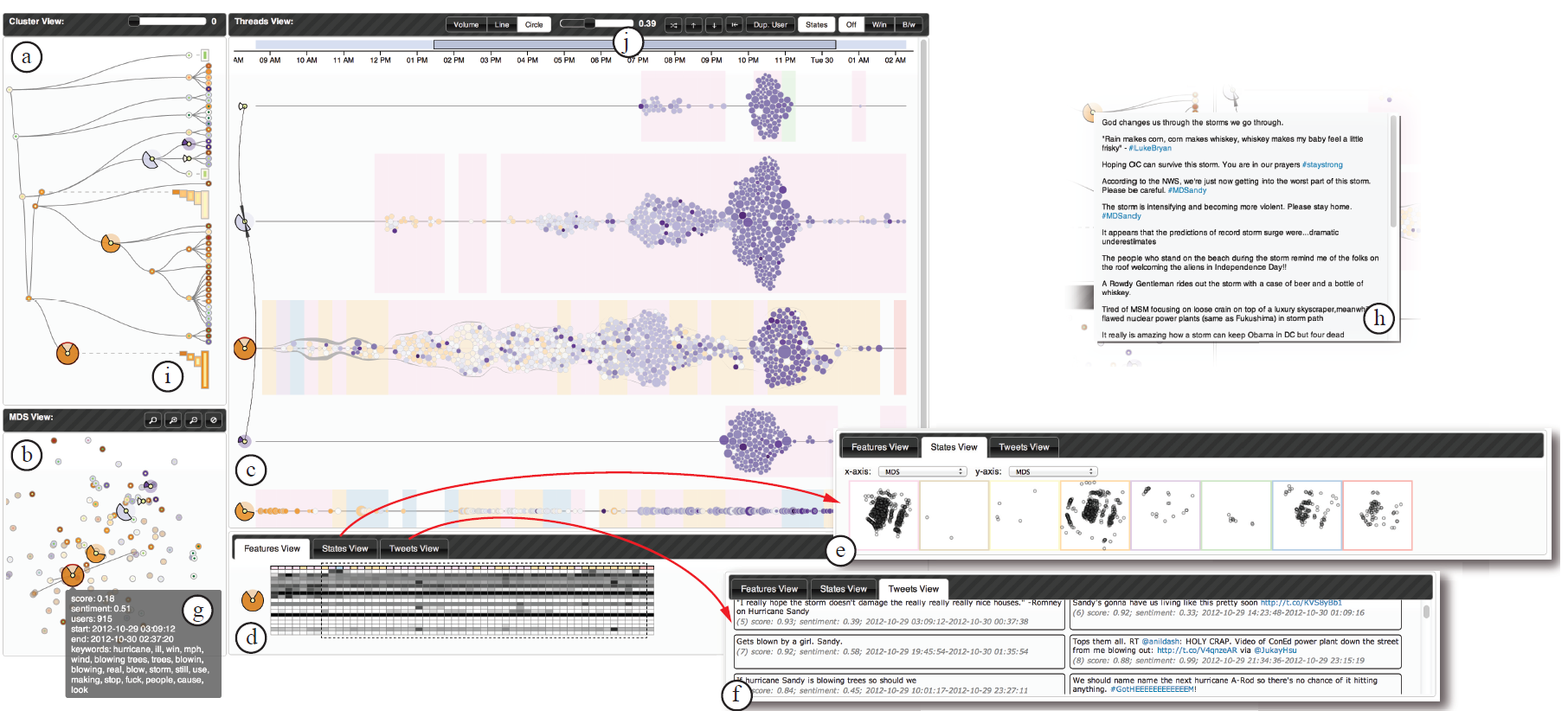} 
	\caption{The FluxFlow system visualizes the hidden states inside the probabilistic model}
\label{fig:fluxflow}
\end{figure*}

The formalism of probabilistic graphical models provides a unifying
framework for capturing complex dependencies among random
variables, and building large-scale multivariate statistical models.
As the names implies,they come with a natural graphical characterization where each nodes encodes a variable, link encodes correlation. 
Visualization for these graphs in 
\cite{chiang2005visualizing, zapata1999visualization} usually provide direct rendering to the graph, highlighting 
the model architecture, transition probabilities, and emission probabilities.
These visualization systems are widely employed in different graphical model inference software, including
Bayes Net Toolbox, Hugin Expert, BayesBuilder, WinMine, BayesianLab, Netica, MSBNx, Analytica, GeNIe/SMILE surveyed in \cite{murphy2001bayes}.

Graphical  Model  usually relies on latent variable to capture hidden structure behind the data and increase expressive power. Techniques for visualizing the latent variable usually exploits its relation between class labels and instance features as well as its transition pattern. The Fluxflow \cite{zhao2014fluxflow}  is a good example.
The hidden state transitions shown as the background
of thread timeline views in Fig. \ref{fig:fluxflow} (j) can reveal the internal stage of
OCCRF. By comparing the state transition patterns across threads, the
user is able to obtain more knowledge about how the model relies on
these states, i.e., user community sub-structures, to perform anomaly
detection.
To further look into the state variables, the analyst can
leverage the features view which summarizes the temporal variations
of feature vectors described with a heatmap-like visualization as shown in Fig. \ref{fig:fluxflow} (d).
From a different perspective of viewing these abstract state variables
, the states view indicates how states are tied to tweet
users by displaying the MDS projections of all users from the highdimensional
feature space as in Fig. \ref{fig:fluxflow} (e). The distributions of users in
these charts can be viewed as signatures of the states characterizing the
features, which helps the analyst understand what each of the abstract
variables might mean.

\section{Discussion and Conclusion}
In this paper, we attempt to survery recent researches and real world systeams integrating the wisdom in two fields towards more effective and efficient data analytics.
We provide a taxonomy of Integrating Information Visualization techniques into Data Mining by identifying three main stages of typical process of data mining, the preliminery analysis of data, the model construction, and the model  evaluation
 and study how each stage can benefit from information visualization.
In reflection, we would say we are satisfied with the amount and scope of literature covered as well as the taxonomy we proposed for the field. If given more time, we would develop a more comprehensive framework for applying such InfoVis method to data mining and systematically discuss the impact of integration.

Future work will involve the providing a more generalizable visualization technique to capture the some unifying concept of data mining and information to 
facilitate tighter and deeper integration for the two fields.
To facilitate the application of these techniques to real world problems, 
method emphasize on the some challenges posed real world data such as 
scalability, sparsity, data noise is another important research direction.
Concrete integrated systems to incorporate the visualization for different types of data mining approach that can efficiently assist the 
data management and mining for both relational data as well as unstructed data are also very important for practical uses.

\bibliographystyle{abbrv} 
\bibliography{VisDMbib}  

\end{document}